\documentclass[journal]{IEEEtran}

\usepackage{mathrsfs}
\usepackage{pifont}
\usepackage{bbding}
\usepackage{amsmath,epsfig}
\usepackage{stmaryrd}
\usepackage{amssymb}
\usepackage{amsfonts}
\usepackage{epic}
\usepackage{graphicx}
\usepackage{subfigure}
\usepackage{curves}
\usepackage{enumerate}
\usepackage{algorithm}
\usepackage{algpseudocode}

\usepackage{stfloats}
\usepackage{latexsym}
\usepackage{epstopdf}
\usepackage{epic}
\usepackage{multirow}
\usepackage{stfloats}
\usepackage{bm}

\usepackage{booktabs}
\usepackage{multirow}                
\usepackage{amsmath}
\usepackage{xcolor}

\begin{document}

\newcommand{\reffig}[1]{Fig.~\ref{#1}}

\newtheorem{definition}{Definition}
\newtheorem{theorem}{Theorem}
\newtheorem{corollary}{Corollary}
\newtheorem{lemma}{Lemma}

\newenvironment{sequation}{\begin{equation}}{\end{equation}}
\newcounter{TempEqCnt}

\renewcommand{\algorithmicrequire}{\textbf{Input:}}   
\renewcommand{\algorithmicensure}{\textbf{Output:}}  

\title{Position-Based Interference Elimination for High Mobility OFDM Channel Estimation in Multi-cell Systems}

\author{Xiang Ren, Wen Chen, \emph{Senior Member, IEEE}, Bo Gong, Qibo Qin, and Lin Gui, \emph{Member, IEEE}
\thanks{Copyright (c) 2015 IEEE. Personal use of this material is permitted. However, permission to use this material for any other purposes must be obtained from the IEEE by sending a request to pubs-permissions@ieee.org.

The authors are with Department of Electronic Engineering,
Shanghai Jiao Tong University, China (e-mail: \{renx,  wenchen, gongbo, qinqibo, guilin\}@sjtu.edu.cn).

This work is supported in part by NSFC under Grant \#61671294, in part by Guangxi NSF key project under Grant
\#2015GXNSFDA139037, in part by Shanghai Key Fundamental Project under Grant \#16JC1402900, and in part by the National Natural Science Foundation of China \#61471236.
}
}

\maketitle

\begin{abstract}

Orthogonal frequency-division multiplexing (OFDM) and multi-cell architecture are widely adopted in current high speed train (HST) systems for providing high data rate wireless communications. In this paper, a typical multi-antenna OFDM HST communication system with multi-cell architecture is considered, where the inter-carrier interference (ICI) caused by high mobility and multi-cell interference (MCI) are both taken into consideration.
By exploiting the train position information, a new position-based interference elimination method is proposed to eliminate both the MCI and ICI for a general basis expansion model (BEM).
We show that the MCI and ICI can be completely eliminated by the proposed method to get the ICI-free pilots at each receive antenna. In addition, for the considered multi-cell HST system, we develop a low-complexity compressed channel estimation method and consider the optimal pilot pattern design.
Both the proposed interference elimination method and the optimal pilot pattern are robust to the train speed and position, as well as the multi-cell multi-antenna system.
Simulation results demonstrate the benefits and robustness of the proposed method  in the multi-cell HST system.

\end{abstract}


\section{Introduction}
High speed trains (HST)  have been developing rapidly around the world and attract  lots of attention \cite{Liu10}-\cite{Karimi12}.
Orthogonal frequency-division multiplexing (OFDM) \cite{Wu1}, \cite{Wu3} and multi-cell architecture, e.g., the microcellular system \cite{Wang12} and the distributed antenna system (DAS) \cite{Karimi12}, are widely adopted in HST wireless communication systems for providing high data rate.
However, in high mobility OFDM systems, the spreading in time and frequency will destroy the orthogonality among subcarriers and introduces inter-carrier interference (ICI). 
In addition, in multi-cell architectures,
the multi-cell interference (MCI) caused by the cells using the same frequency is inevitable~\cite{Wang12}, especially at the cell edge of the adjacent cells.
The ICI and MCI will directly reduce the channel estimation accuracy resulting in degraded system performance.


Channel estimation is a non-trivial problem  in high mobility OFDM systems. 
Many channel estimation techniques have been proposed based on different channel characteristics \cite{Mostofi05}-\cite{Ren15}.
The methods in \cite{Mostofi05} and \cite{Kwak10} are based on a piece-wise linear channel model, which assumes that the channel varies with time linearly in one or more OFDM symbols.
The works \cite{Tang07}-\cite{Ma03} resort to estimating the equivalent discrete-time channel taps modeled by basis expansion models (BEM).
%
In \cite{Bajwa10}-\cite{Ren15}, the authors considered compressed sensing (CS) based channel estimation methods to utilize the inherent channel sparsity.
The works \cite{Bajwa10}-\cite{Gui14} propose several CS-based estimation methods without considering the effect of a large Doppler shift.
In \cite{Ren13} and \cite{Ren15ICC}, CS-based channel estimation methods with designed pilot are developed for  OFDM systems over high mobility channels.
In \cite{Ren14} and \cite{Ren15}, two position-based compressed channel estimation methods are developed for HST systems, where the train position information is utilized to improve the estimation performance and combat the ICI.
Moveover, many other applications highly depend on accurate channel estimation schemes, such as green communication and wireless power transfer \cite{Wu2}-\cite{Wu5}.

To combat the ICI effect in high mobility systems, many ICI mitigation methods have been developed \cite{Ren14}, \cite{Ren15}, \cite{Campos13}-\cite{Cheng13}.
The authors of \cite{Ren14}, \cite{Campos13}-\cite{Simon12} propose several ICI mitigation methods based on iterative process, which incur high complexity for a large Doppler shift.
The work \cite{Ren15} proposes a position-based ICI elimination method for the single-input multiple-output (SIMO) OFDM HST system, where ICI-free pilots can be obtained for the complex exponential BEM (CE-BEM) by exploiting the train position information.
In \cite{Cheng13}, an ICI-free pilot structure is proposed for OFDM systems in the CE-BEM, where a large number of guard pilots is needed resulting a low spectrum efficiency.
Both \cite{Ren15} and \cite{Cheng13} are only designed for the CE-BEM and will suffer from residual ICI for other BEMs, resulting in degraded system performance.
These aforementioned works seldom consider the multi-cell system, which, however, is widely adopted in current HST systems to provide high data rate services.

In multi-cell HST systems,
base stations (BS) are generally evenly allocated along the railway to communicate with the mobile users in HSTs via a relay station (RS) installed on the train \cite{Wang12}, \cite{Karimi12}. 
This architecture divides the railway into many small cells and can provide high data rate services by shortening the transmission distance between the transmitter and the receivers.
Generally, there exists a overlap between each two adjacent cells, which incurs the MCI by cells using the same frequency.
Due to the high speed of train and the small cell size, the train will move across the cell edge frequently and the MCI significantly degrades the overall system performance.
To solve this problem, one method is to use the specific antenna to execute handover with the target BS while other antennas communicating with the serving BS \cite{Ren14}, \cite{Tian12}.
This method needs additional costs and incurs high complexity for antenna selections, especially for large-scale antenna systems.
Another technique is frequency reuse \cite{Karimi12}, \cite{Zhu11}, which considers that the adjacent cells use different frequencies (typical frequency reuse factor is $1/3$ \cite{Zhu11}). This method can effectively eliminate the MCI, which, however, highly reduces the spectrum efficiency since the total frequency is divided into several subsets and each cell uses one subset.
Therefore, MCI mitigation in multi-cell HST systems becomes a severe problem that must be considered.

In this work, different to our previous works \cite{Ren14} and \cite{Ren15} based on single-cell scenarios, we consider a more practical multi-cell multi-antenna HST communication system, where the ICI caused by high mobility and the MCI at the cell edge are both taken into consideration. 
Note that both \cite{Ren14} and \cite{Ren15} cannot be directly applied to multi-cell systems due to the inevitable MCI.
In addition, different from \cite{Ren15} that only considers the CE-BEM, in this work, we consider a general BEM based channel model.

We first exploit the position information of the high mobility channel modeled by a general BEM, and propose a simplified position-based channel model.
Next, with the proposed position-based MCI elimination method, we show that the
signals transmitted from different cells can be separated at the receive antenna corresponding to their different Doppler shifts.
Then, a new position-based ICI elimination method is proposed for a general BEM to get the ICI-free pilots at the receive antenna.
In specific, an example in the generalized complex exponential BEM (GCE-BEM) is given to verify the effectiveness of the proposed method.
In contrast to the methods in \cite{Ren14} and \cite{Ren15} that need additional guard pilots and complexity to combat the MCI, the proposed method can eliminate both the MCI and ICI without the help of guard pilots, which highly improves the spectral efficiency.
In addition, different from the method in \cite{Ren15} which can only get the ICI-free pilots for the CE-BEM and will suffer from residual ICI for other BEMs, the proposed method can obtain the ICI-free pilots for a general BEM.
After that, a low-complexity compressed channel estimation method with optimal pilot pattern design is developed for the considered multi-cell HST system.
Particularly, the optimal pilot pattern is independent of the train speed and position, the number of antennas, and the number of cells.
Simulation results verify the benefits of the proposed scheme in the considered HST system.
Moreover, compared to the method in \cite{Ren15} whose the system performance is significantly influenced by the multi-cell architecture and the channel model, it is shown that the proposed method is robust to the MCI and BEMs.

The rest of this paper is organized as follows. Section II describes the considered system model and introduces the channel model based on the BEM.  In Section III, we exploit the position information and then propose the position-based MCI and ICI elimination method.
A low-complexity compressed channel estimation method with the optimal pilot pattern is developed for the considered system in Section IV, where the complexity and the summary of the proposed method are also given.
Simulation results are presented in Section V. At last, Section VI provides the conclusions.

$Notations$:
$\left\|\cdot\right\| _{\ell _0 }$ indicates the number of nonzero entries in a matrix or vector, and $\left\|\cdot\right\| _{\ell _2 }$ is the Euclidean norm.
The superscripts $(\cdot)^T$, $(\cdot)^H$, $(\cdot)^{-1}$ represent transpose, complex conjugate transpose, inverse, respectively.
$\lceil\cdot\rceil$ and $\lfloor\cdot\rfloor$ indicate round up and round down operators, respectively.
$\star$ stands for a dot product operator,  and ${\bf A}=\text{diag}\{{\bf a}\}$ stands for a diagonal matrix $\bf A$ with a vector $\bf a$ on its main diagonal.
 In addition, we denote the $K\times K$ all-one matrix as ${\bf 1}_K$,  the all-zero matrix as $\bf 0$, and the the identity matrix as ${\bf I}$.
${\bf X} (:,{\bf w})$ denotes a submatrix of the matrix ${\bf X}$ with column indices $\bf w$ and all rows,
and ${\bf Y} ({\bf p},:)$ denotes a submatrix of ${\bf Y}$ with row indices $\bf p$ and all columns.
Finally, $\mathbb{C}^{M\times N}$ and $\mathbb{R}^{M\times N}$ stand for the set of $M\times N$ matrices in complex field and in real field, respectively.

\section{System Model}
\subsection{Multi-cell HST Communication System}

\begin{figure*}[t!]
  \centering
  \includegraphics[width= 6.2in]{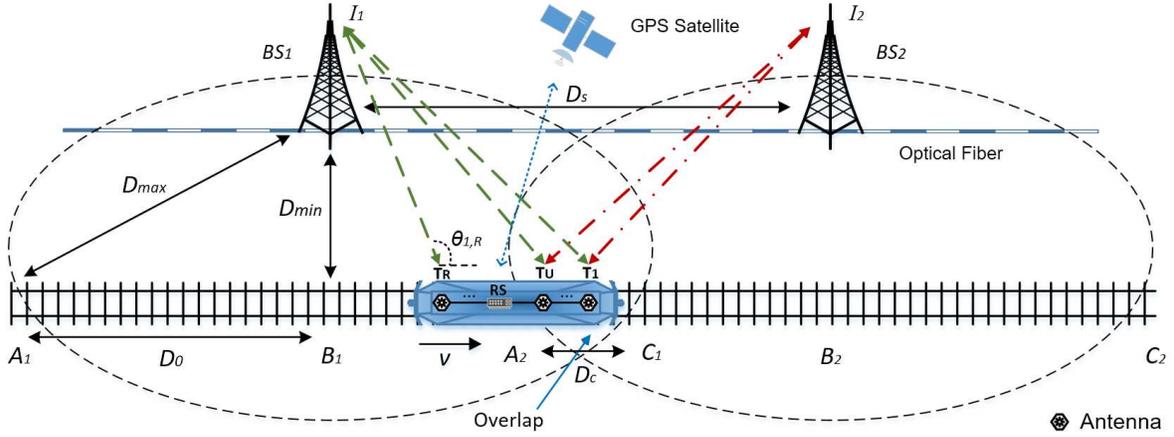}\\
  \caption{The structure of a multi-cell HST communication system.}\label{sys}
\end{figure*}

We consider a well-recognized HST wireless communication system with the multi-cell network architecture \cite{Wang12, Karimi12}, as illustrated in \reffig{sys}.
The RS has multiple antennas on the top of the train to communicate with the BS, and multiple antennas are distributed in carriages to communicate with the mobile users.
The BSs are evenly allocated along the railway and connected with optical fibers, dividing the system into a number of small cells. For each cell, we assume that it has a BS with one antenna, has the same coverage range, and uses the same frequency \cite{WINNER07}.\footnote{One may also consider that the BS is equipped with multiple antennas. In such case, some advanced techniques, e.g., precoding and space-time coding \cite{Hlawatsch11}, can be utilized to obtain the multiplexing gain and diversity gain. In addition, some cancellation techniques in \cite{Hlawatsch11} can be adopted to combat the incurred inter-antenna interference (IAI).}

The HST is equipped with a global positioning system (GPS) to acquire the train position and speed information, which are vital for train control systems to guarantee the train-running safety.
Typically, the GPS provides an average accuracy better than 3.6m in the open space \cite{Liu11}. However, its accuracy may be harmed by several factors, e.g., atmospheric effects, terrains, and environments. For the environments surrounded with trees and/or buildings, the location error of the GPS  may reach up to 11--17m \cite{Liu11}. In such case, higher accuracy is attainable by using GPS in combination with other positioning systems, such as transponders, track circuits, axle counters, and so on \cite{Pascoe09}.
In this work, we consider the train runs in an open plain and assume that there exists a strong line-of-sight (LOS) path between each BS and each receive antenna on the HST.{\footnote{In practice, there are some non-LOS scenarios exist in HST systems due to the shelters of trees, mountains or buildings, and tunnels as well. However, the newly-built HST routes are usually designed with gentler curves, shallower grooves, wider tunnels, and the BSs are allocated along the railway with a small distance and a high antenna height, yielding a clear or open space with a strong LOS path \cite{Liu10}. These make the LOS scenarios more dominant in HST systems. }} 
In addition, as the GPS provides satisfactory accuracy in the open space, we also assume that the GPS estimates the train position and speed information perfectly and sends them to the HST without time delay.
Detailed impacts of the GPS location error on  the proposed method will be discussed in Section IV-E.

In \reffig{sys}, assume the HST moves at a constant speed~$v$.
Denote $D_{max}$ as the coverage of the BS, i.e.,  the distance between $A_1$ and $BS_1$, $D_{min}$ as the minimum distance between the BS and the railway, i.e., $B_1$ to $BS_1$, $D_0$ as the distance between $A_1$ and $B_1$, $D_s$ as the distance between $BS_1$ and $BS_2$, and $D_c$ as the length of the overlap $A_2-C_1$.
Let $\{T_r\}^R_{r=1}$ denote the receive antennas of the RS which are evenly located on the top of the HST,
and  $\{I_t\}^{T=2}_{t=1}$ denote the transmit antenna of $BS_t$.
For each cell, define $\alpha_{t,r}\in[0, 2D_0]$ as the distance between $T_r$ and $A_t$, where $\alpha_{t,r} = 0$ at $A_t$ and $\alpha_{t,r} = 2D_0$ at $C_t$,  and $\theta_{t,r}$ as the angle between the direction of the train speed and $BS_t$ to $T_{r}$.
In general, $\theta_{t,r}$ can be directly calculated by the HST's position information $\alpha_{t,r}$ (supported by the GPS) and its relative position to $BS_t$.
Note that, since the train moves along the railway, the geographical locations of BSs (usually installed along the railway track) can be conveniently obtained and pre-stored at the HST, e.g., using track map \cite{Pascoe09}.
When the $r$-th receive antenna moves at a certain position $\alpha_{t,r}$, it suffers from a Doppler shift from $BS_t$ as $f_{t,r} = \frac{v}{c}\cdot f_c \cos\theta_{t,r}$,  where  $f_c$ is the carrier frequency, and $c$ is the speed of light.
In \reffig{sys}, considering a case of the HST moving from $A_2$ to $C_1$, as we consider that $BS_1$ and $BS_2$ use the same frequency, it is easy to find that the $U$ ($U\leq R$) receive antennas in the overlap receive both the signals from $BS_1$ and $BS_2$, resulting in the MCI.

\subsection{MIMO-OFDM System}
Considering that $U$ receive antennas are at the overlap $A_2-C_1$,
 it can be treated as a multiple-input multiple-output (MIMO) OFDM system with 2 transmit antennas,   $U$ receive antennas, and $K$ subcarriers.
Denote $X_t(k)$ as the signal transmitted by $BS_t$ over the $k$-th subcarrier during one OFDM symbol, where $k = 0, 1, ..., K-1$ and $t = 1,2$.
After performing the inverse discrete Fourier transform (IDFT) operation at each BS,
cyclic prefix (CP) is inserted into the transmitted signal to avoid the intersymbol interference (ISI).
At each receive antenna, the received signal is demodulated by the discrete Fourier transform (DFT) operation after removing the CP.
For the $r$-th receive antenna, the received frequency domain signal ${\bf{y}}_r = [Y_r(0),Y_r(1),...,Y_r(K-1)]^T$ is expressed as
\begin{align}
  {\bf y}_r = \sum^{T=2}_{t=1}{\bf H}_{t,r}{\bf x}_t+{\bf n}_r,\label{eq1}
\end{align}
where ${\bf{H}}_{t,r}$ is the frequency domain channel matrix from $BS_t$ to the $r$-th receive antenna,
${\bf{x}}_t= [X_t(0),X_t(1),...,X_t(K-1)]^T$ is the signal vector transmitted by $BS_t$,
${\bf{n}}_r = [N_r(0),N_r(1),...,N_r(K-1)]^T$ is the noise vector at the $r$-th receive antenna, and $N_r(k)$ is the additive white Gaussian noise (AWGN) with a zero mean and $\sigma^2_\varepsilon$ variance over the $k$-th subcarrier.

For the time-invariant channel, ${\bf H}_{t,r}$ will be a diagonal matrix and the received signal is free of the ICI.
However, for high mobility channels, ${\bf H}_{t,r}$ becomes a full matrix, resulting in the ICI.
Denote ${\bf{H}}^{\rm free}_{t,r} \triangleq {\text{diag}}\{[H_{t,r}(0,0),H_{t,r}(1,1),...,H_{t,r}(K-1,K-1)]\}$ as the ICI-free channel matrix of ${\bf H}_{t,r}$, and ${\bf{H}}^{\rm ICI}_{t,r} \triangleq {\bf{H}}_{t,r} - {\bf{H}}^{\rm free}_{t,r}$ denotes the ICI part.
Then,  (\ref{eq1}) can be rewritten as
\begin{align}
{\bf y }_r  &=\sum^{T=2}_{t=1}{\bf H}^{\rm{free}}_{t,r}{\bf x}_t + {\bf H}^{\rm{ICI}}_{t,r}{\bf{x}}_t + {\bf{n}}_r, \label{eq2}\\
& = \underbrace{{\bf H}^{\rm{free}}_{1,r}{\bf x}_1 + {\bf H}^{\rm{ICI}}_{1,r}{\bf{x}}_1}_{{\bf Y}_{1,r}} +\underbrace{ {\bf H}^{\rm{free}}_{2,r}{\bf x}_2 + {\bf H}^{\rm{ICI}}_{2,r}{\bf{x}}_2}_{{\bf Y}_{2,r}}+ {\bf{n}}_r,
\end{align}
where  ${\bf Y}_{1,r}$ denotes the signal transmitted from $BS_1$ to the $r$-th receive antenna, and ${\bf Y}_{2,r}$ denotes the signal transmitted from $BS_2$. It is easy to find that ${\bf y }_r$ is a sum of the signals from the two neighbouring BSs.

\subsection{Channel Model based on BEM}
In this paper,  the widely used BEM is adopted to model the high mobility channel between BSs and the receive antennas of the RS \cite{Tang07}.
Assume there are $L$ multi-paths between each transmit antenna and each receive antenna.
Define ${{\bf{h}}}_{t,r}(l)= [h_{t,r}(0,l), h_{t,r}(1,l), ..., h_{t,r}(K-1,l)]^T\in\mathbb{C}^{K\times1}$, 
where $h_{t,r} (\tilde{k},l)$ denotes the $l$-th channel tap between $BS_t$ and the $r$-th receive antenna at the $\tilde{k}$-th time instant, $ l = 0,1,..., L-1$, and ${h_{t,r}(\tilde{k},l)} = 0$ for $l>L-1$.
Then,  ${\bf{h}}_{t,r}(l)$ can be represented as
\begin{align}
  {{\bf{h}}}_{t,r}(l)= {\bf B}{\bf c}_{t,r}(l) + {\boldsymbol {\epsilon}}_{t,r}(l),\label{eq5}
\end{align}
where ${\bf B} = [{\bf b}_{0},...,{\bf b}_q,...,{\bf b}_{Q}] \in\mathbb{C}^{K\times (Q+1)}$ is the BEM basis matrix, ${\bf b}_q = [{{b}_q(0),{b}_q(1) ,...,{ b}_q(K-1)}]^T\in\mathbb{C}^{K\times 1}$ is the $q$-th  ($q = 0,1,...,Q$) basis function vector, ${\bf c}_{t,r}(l) = [c_{t,r}(0,l),c_{t,r}(1,l),...,c_{t,r}(Q,l)]^T$ collects the $Q+1$ BEM coefficients of the $l$-th channel tap,
$Q = 2\lceil f_{{\max}} T_d\rceil$ is the BEM order with the maximum Doppler shift $f_{\max}$ and the packet duration $T_d$, and ${\bm \epsilon}_{t,r}(l) = [\epsilon_{t,r}(0,l),\epsilon_{t,r}(1,l),...,\epsilon_{t,r}(K-1,l)]^T$ represents the BEM modeling error.

According to (\ref{eq5}), we have
 \begin{align}
 h_{t,r} (\tilde{k},l) = \sum^Q_{q=0}{b}_q(\tilde{k})c_{t,r}(q,l) + \epsilon_{t,r}(\tilde{k},l), \label{ap2}
\end{align}
where ${ b}_q(\tilde{k})$ is the $\tilde{k}$-th entry of ${\bf b}_q$, and $c_{t,r}(q,l)$ is the $q$-th entry of ${\bf c}_{t,r}(l)$.

Denote $\tilde{\mathbf{H}}_{t,r}$ as the time domain channel matrix between $BS_t$ and the $r$-th receive antenna which exhibits a pseudo-circular structure as  \cite{Tang07}
\begin{align}
\tilde{{\bf H}}_{t,r}(\tilde{k},d) = h_{t,r}(\tilde{k},|\tilde{k}-d|_K),~~\tilde{k},d\in[0,K-1],\label{eq003}
\end{align}
where $\tilde{{\bf H}}_{t,r}(\tilde{k},d)$ indicates the $(\tilde{k},d)$-th entry of $\tilde{{\bf H}}_{t,r}$, and $|\cdot|_K$ denotes a mod $K$ operator.
Inserting (\ref{ap2}) into (\ref{eq003}),
$\tilde{\mathbf{H}}_{t,r}$ can be expressed as
\begin{equation}
 \tilde{\mathbf{H}}_{t,r} = \sum^Q_{q=0} \tilde{{\bf D}}_q {\bf C}_{t,r,q} + \tilde{\boldsymbol{\xi}}_{t,r},\label{ap3}
\end{equation}
where $\tilde{{\bf D}}_q = \text{diag}\{{\bf b}_q\}$, ${\bf C}_{t,r,q}$ is a $K\times K$ circulant matrix with $\bar{{\bf c}}_{t,r,q} \triangleq [c_{t,r}(q,0),c_{t,r}(q,1),...,c_{t,r}(q,L-1), {\bf 0}_{1\times(K-L)}]^T$ as its first column, ${\bf 0}_{1\times(K-L)}$ indicates a $1\times (K-L)$ all-zero vector, and $\tilde{\boldsymbol{\xi}}_{t,r}$ is the BEM modeling error matrix in the time domain.

Then, we can write the frequency domain channel matrix between $BS_t$ and the $r$-th receive antenna as
\begin{align}
{\bf H}_{t,r}  &={\bf F}\tilde{\mathbf{H}}_{t,r}{\bf F}^H + {\boldsymbol{\xi}}_{t,r},\\
&=  \sum^Q_{q=0} {{\bf D}}_q {\bf \Delta}_{t,r,q} + {\boldsymbol{\xi}}_{t,r}, \label{eqBEM}
\end{align}
where ${\bf D}_q = {\bf F}{\text{diag}}\{{\bf b}_q\}{\bf F}^H$,  ${\bf \Delta}_{t,r,q} = {\text{diag}}\{{\bf F}_L{\bf c}_{t,r,q}\}$ is a diagonal matrix with $ {\bf c}_{t,r,q} \triangleq [c_{t,r}(q,0),...,c_{t,r}(q,L-1)]^T$, 
$\bf F$ is the DFT matrix, ${\bf F}_L$ is the first $L$ columns of $\sqrt{K}{\bf F}$, and ${\boldsymbol{\xi}}_{t,r}$ is the BEM modeling error matrix in the frequency domain.
In the following, since we mainly focus on  the MCI and ICI elimination and the channel estimation, the BEM modeling error ${\boldsymbol{\xi}}_{t,r}$ will be omitted for the convenience of illustration. In fact, ignoring  ${\boldsymbol{\xi}}_{t,r}$  will not affect the detailed expressions and the conclusions of the proposed method.

\subsection{Channel Estimation based on BEM}
From now on, we can describe our system based on the BEM.
Substituting (\ref{eqBEM}) into (\ref{eq1}),  we obtain
\begin{align}
  {\bf y}_r
  & = \sum^{T}_{t=1} \sum^Q_{q=0} {{\bf D}}_q {\bf \Delta}_{t,r,q}{\bf x}_t+{\bf n}_r.\label{eq9}
\end{align}

Assume that $BS_t$ transmits $P$ ($P<K$) pilots at the subcarrier pattern ${\bf w}_t =[w_{t,1},w_{t,2},...,w_{t,P}]$ and  the data are transmitted at the subcarrier pattern ${\bf d}_t$.
The received pilots at the $r$-th receive antenna with ${\bf w}_t$ can be represented as
\begin{align}\
  {\bf y}_{r}({\bf w}_t) =   &\underbrace{\sum^Q_{q=0}{\bf D}_q({\bf w}_t,{\bf w}_t){\bf \Delta}_{t,r,q}({\bf w}_t,{\bf w}_t){\bf x}_t({\bf w}_t)}_{{\bf \Omega}_{t,r}}\nonumber\\
  &~~+ \underbrace{ \sum^Q_{q=0}{\bf D}_q({\bf w}_t,{\bf d}_t){\bf \Delta}_{t,r,q}({\bf d}_t,{\bf d}_t){\bf x}_t({\bf d}_t)}_{{\bf \Upsilon}_{t,r}} \nonumber\\ &~~~~+ \underbrace{\sum^{T}_{\nu=1,\nu\neq t} \sum^Q_{q=0} {{\bf D}}_q({\bf w}_t,{\bf k}_\nu) {\bf \Delta}_{\nu,r,q}({\bf k}_\nu,{\bf k}_\nu){\bf x}_\nu({\bf k}_\nu)}_{{\bf A}_{t,r}}\nonumber\\
   &~~~~~~+{\bf n}_r({\bf w}_t), \label{eq26}
\end{align}
where  ${\bf k}_\nu = {\bf w}_\nu\cup{\bf d}_\nu$ denotes the union set of the pilot and data pattern of  $BS_\nu$, and $\nu\neq t$.
In (\ref{eq26}), for the desired pilot ${\bf \Omega}_{t,r}$, we first decouple the ICI introduced by the data transmitted from $BS_t$ and put it in the term ${\bf \Upsilon}_{t,r}$, and then decouple the MCI caused by the signals transmitted from other BSs and put them in the term ${\bf A}_{t,r}$.

For time-varying channels, it can be found that the desired ${\bf \Omega}_{t,r}$ is distorted by ${\bf \Upsilon}_{t,r}$ due to ${\bf \Upsilon}_{t,r}\neq {\bf 0}$.
In addition, for the considered system in \reffig{sys}, as we assume that $BS_1$ and $BS_2$ use the same frequency, it is easy to have  ${\bf A}_{t,r} = {\bf 0}$ for the $r$-th receive antenna moving from $A_1$ to $A_2$, whereas  ${\bf A}_{t,r} \neq {\bf 0}$ for it moving into the overlap $A_2-C_1$.
When the train passes the overlap at a high speed, ${\bf \Upsilon}_{t,r}$ and ${\bf A}_{t,r}$ will directly degrade the channel estimation performance.
In addition, due to the high train speed and the small cell size, the train will pass the overlaps frequently and ${\bf A}_{t,r}$ is  inevitable.
Therefore, MCI and ICI eliminations are necessary for multi-cell HST systems.

\section{Position-based MCI and ICI Elimination}

In this section, we first exploit the position information of the BEM and propose a position-based channel model. Then, for a general BEM channel model, we show that both the MCI and the ICI can be eliminated at each receive antenna by exploiting the position information. In addition, an example in the GCE-BEM is given for better clarification.

\subsection{Position Information of BEM}
In this subsection, we first introduce a definition and a lemma based on the BEM given in (\ref{eqBEM}).

\begin{definition}[Channel Sparsity \cite{Bajwa10}]\label{df1}
For a wireless channel based on the BEM, the channel coefficients that contribute significant powers are called as the dominant coefficients, i.e., $|{c_{t,r}(q,l)}|^2 > \gamma$, where $\gamma$ is a pre-fixed threshold.
These dominant coefficients reflect the major properties of the channel while other coefficients with minor powers can be neglected.
The channel ${\bf H}_{t,r}$ is $S$-sparse if  $S = \left\| {{\bf{c}}}_{t,r} \right\|_{\ell _0 }  \ll L(Q+1)$, where ${\bf c}_{t,r} = [{\bf c}_{t,r,0}^T,{\bf c}^T_{t,r,1},...,{\bf c}^T_{t,r,Q}]^T\in\mathbb{C}^{L(Q+1)\times 1} $ collects all the BEM coefficients of ${\bf H}_{t,r}$.
\end{definition}

\begin{lemma}[Position-based channel sparsity \cite{Ren15}] For an HST system, if the high mobility channel  between $BS_t$ ($t=1,2$) and the $r$-th ($r=1,2,...,R$) receive antenna is $S$-sparse, then ${\bf H}_{t,r}$ is $S$-sparse at any given position and its dominant BEM coefficients only exist in ${\bf c}^*_{t,r}$, i.e.,
\begin{align}
{\bf{c}}^*_{t,r} &= {\bf{c}}_{t,r,q|q=q^*_{t,r}}\\
&=
\left[\begin{matrix}
   c_{t,r}(q^*_{t,r},0), c_{t,r}(q^*_{t,r},1), ..., c_{t,r}(q^*_{t,r},L-1)
\end{matrix}\right]^T,
\end{align}
where $q^*_{t,r}\in\{0,1,...,Q\}$ is called as the dominant index of ${\bf H}_{t,r}$ corresponding to $f_{t,r}$.
\begin{proof}
Please see \cite{Ren15}.
\end{proof}
\end{lemma}

Next, the following theorem is given to explore the position information of the considered multi-cell HST system.

\begin{theorem} For a multi-cell HST system with any given train position, if the high mobility channel between $BS_t$ ($t=1,2$) and the $r$-th ($r=1,2,...,R$) receive antenna is $S$-sparse, it can be modeled with its dominant coefficients ${\bf c}^*_{t,r}$ and the dominant basis function ${\bf D}^*_{t,r} = {\bf D}_{q|q={q}^*_{t,r}}$. In specific,
\begin{align}
{\bf H}_{t,r}  = \left\{\begin{matrix}
  &{\bf D}^*_{t,r}{\bf \Delta}^*_{t,r},~~&\alpha_{t,r}\in\left[0,2D_0\right],\\
  &{\bf 0},~~&\alpha_{t,r}<0~~ \text{or}~~ \alpha_{t,r}>2D_0.
\end{matrix}\right.
\end{align}
 where ${\bf \Delta}^*_{t,r} = {\text{diag}}\{{\bf F}_L{\bf c}^*_{t,r}\}$, $\alpha_{t,r}\in\left[0,2D_0\right]$ denotes that the $r$-th receiver is in the cell of $BS_t$, and $\alpha_{t,r}<0~~ \text{or}~~ \alpha_{t,r}>2D_0$ denotes that the $r$-th receive is out of the cell.
\begin{proof}
Let us consider the multi-cell HST system in \reffig{sys}.  Assume that
the HST moves into the cell of $BS_t$ and the $r$-th receive antenna moves at a certain position $\alpha_{t,r}$ with the Doppler shift $f_{t,r}$.
Then, the  channel between $BS_t$ and
the $r$-th receive antenna can be represented as  (\ref{eqBEM}), where ${\bf H}_{t,r}$ is represented as the sum of the products of the basis matrix ${\bf D}_q$ and the corresponding BEM coefficients ${\bf \Delta}_{t,r,q}$ over all Doppler shifts.
Note that $q = 0,1,...,Q$ correspond to the Doppler shifts from $-f_{\max}$ to $f_{\max}$ in sequence, e.g., $q = 0$ for  $-f_{\max}$ and   $q = Q$ for $f_{\max}$, respectively.

According to Lemma 1, if ${\bf H}_{t,r}$ is $S$-sparse, its dominant coefficients only exist in ${\bf c}^*_{t,r}$ with the index $q^*_{t,r}$ corresponding to $f_{t,r}$,
and the non-dominant ones can be neglected, i.e.,
${\bf c}_{t,r,\bar q} = {\bf 0}$ for $\bar q = 0,1,...,Q$ and $\bar q\neq q^*_{t,r}$. Therefore, we can rewrite (\ref{eqBEM}) as
\begin{align}
{\bf H}_{t,r} 
& = {{\bf D}}_{q^*_{t,r}} {\bf \Delta}_{t,r,{q^*_{t,r}}} + \underbrace{\sum^Q_{\bar q=0, \bar q \neq q^*_{t,r}} {{\bf D}}_{\bar q} \text{diag}\{{\bf F}_{L}{\bf c}_{t,r,\bar q}\}}_{\bf 0}, \label{T2}\\
& ={\bf D}^*_{t,r}{\bf \Delta}^*_{t,r}. \label{T3}
\end{align}
Note that we here assume that all paths  suffer from identical Doppler shifts. This is reasonable since there usually exists a strong LOS propagation path in HST channels \cite{Liu10}.
Therefore, for the $r$-th receive antenna moves at $\alpha_{t,r}$ with $f_{t,r}$, it can be found that all the channel taps of ${\bf H}_{t,r}$ suffer from the same $f_{t,r}$ and correspond to the same index $q^*_{t,r}$ and thus the same ${\bf c}^*_{t,r}$. In this way, ${\bf H}_{t,r}$ can be represented with the dominant parts ${\bf \Delta}^*_{t,r} = {\text{diag}}\{{\bf F}_L{\bf c}^*_{t,r}\}$
 and ${\bf D}^*_{t,r}$  as (\ref{T3}).
The relationships between $q^*_{t,r}$, $\alpha_{t,r}$ and $f_{t,r}$ are given in the following part.

Besides, for $\alpha_{t,r}<0$ and $ \alpha_{t,r}>2D_0$, it is easy to have ${\bf H}_{t,r}  = {\bf 0}$ since the $r$-th receive antenna is out of the coverage range of $BS_t$.
\end{proof}
\end{theorem}

For the $r$-th receive antenna at the cell $BS_t$, the relationship between  $q^*_ {t,r}$ and  $f_{t,r}$ is expressed as
\begin{align}
q^*_{t,r}=\left\{\begin{matrix}
  &\left\lceil{T_d}{f_{t,r}}\right\rceil+\frac{Q}{2},&f_{t,r}\in\left[0,f_{{\max}}\right],\\
  &\left\lfloor{T_d}{f_{t,r}}\right\rfloor+\frac{Q}{2},&f_{t,r}\in\left[-f_{{\max}},0\right).
\end{matrix}\right.\label{q_fd}
\end{align}
In addition, the relationship between $q^*_{t,r}$ and the antenna position $\alpha_{t,r}$ is given as
\begin{align}\small
q^*_{t,r}=\left\{\begin{matrix}
  &\left\lceil{F}\cdot\frac{D_0-\alpha_{t,r}}{\sqrt{{(D_0-\alpha_{t,r})}^2+{D_{{min}}}^2}}\right\rceil+\frac{Q}{2},~&{\alpha_{t,r}}\in[0,D_0],\\ \\
   &\left\lfloor{F}\cdot\frac{D_0-\alpha_{t,r}}{\sqrt{{(D_0-\alpha_{t,r})}^2+{D_{{min}}}^2}}\right\rfloor+\frac{Q}{2},~&{\alpha_{t,r}}\in(D_0,2D_0],
\end{matrix}\right.\label{q_a}
\end{align}
where ${F} =T_d  f_{{\max }} =T_d\frac{v}{c} \cdot f_c$, ${\alpha_{t,r}}\in[0,D_0]$ denotes the $r$-th receive antenna moving from $A_t$ to $B_t$, and ${\alpha_{t,r}}\in(D_0,2D_0]$ denotes the positions from $B_t$ to $C_t$.

According to Theorem 1, when the $r$-th receive antenna moves into the overlap $A_2-C_1$, (\ref{eq9}) can be thus represented as
\begin{align}
  {\bf y}_r
  & = \sum^{T=2}_{t=1} {{\bf D}}^*_{t,r} {\bf \Delta}^*_{t,r}{\bf x}_t+{\bf n}_r,\\
& =  {{\bf D}}^*_{1,r} {\bf \Delta}^*_{1,r}{\bf x}_1+{{\bf D}}^*_{2,r} {\bf \Delta}^*_{2,r}{\bf x}_2+{\bf n}_r.\label{eq14}
\end{align}
Note that we have ${{\bf D}}^*_{1,r}\neq{{\bf D}}^*_{2,r}$ in the considered multi-cell HST system.  For the $r$-th receive antenna in the overlap $A_2-C_1$, it is easy to find that $f_{1,r}\neq f_{2,r}$ ($f_{1,r}<0$ and $ f_{2,r}>0$) due to its different relative positions to $BS_1$ and $BS_2$, and we have $q^*_{1,r}\neq q^*_{2,r}$ and ${{\bf D}}^*_{1,r}\neq{{\bf D}}^*_{2,r}$.
We also have $ {\bf \Delta}^*_{r,1}\neq {\bf \Delta}^*_{r,2}$ due to ${\bf H}_{1,r}$ and ${\bf H}_{2,r}$ are independent. In (\ref{eq14}), it is obvious that the receive antenna receives both the signals from the two adjacent BSs, incurring the MCI.
In addition, since $ {{\bf D}}^*_{t,r}$ is approximately banded for most BEMs, there still exists ICI in (\ref{eq14}).

On the other hand, when the $r$-th receive antenna is out of the overlap $A_2-C_1$, we have
\begin{align}
    {\bf y}_r  =  {{\bf D}}^*_{t,r} {\bf \Delta}^*_{t,r}{\bf x}_t+{\bf n}_r.\label{eq13}
\end{align}
From (\ref{eq13}),  since the receive antenna can only receive the signal transmitted by $BS_t$, it is easy to find that the $r$-th receive antenna is free of the MCI but with the ICI.


\subsection{Position-based MCI and ICI Elimination}

\begin{figure*}[t!]
  \centering
  \includegraphics[width=5.4in]{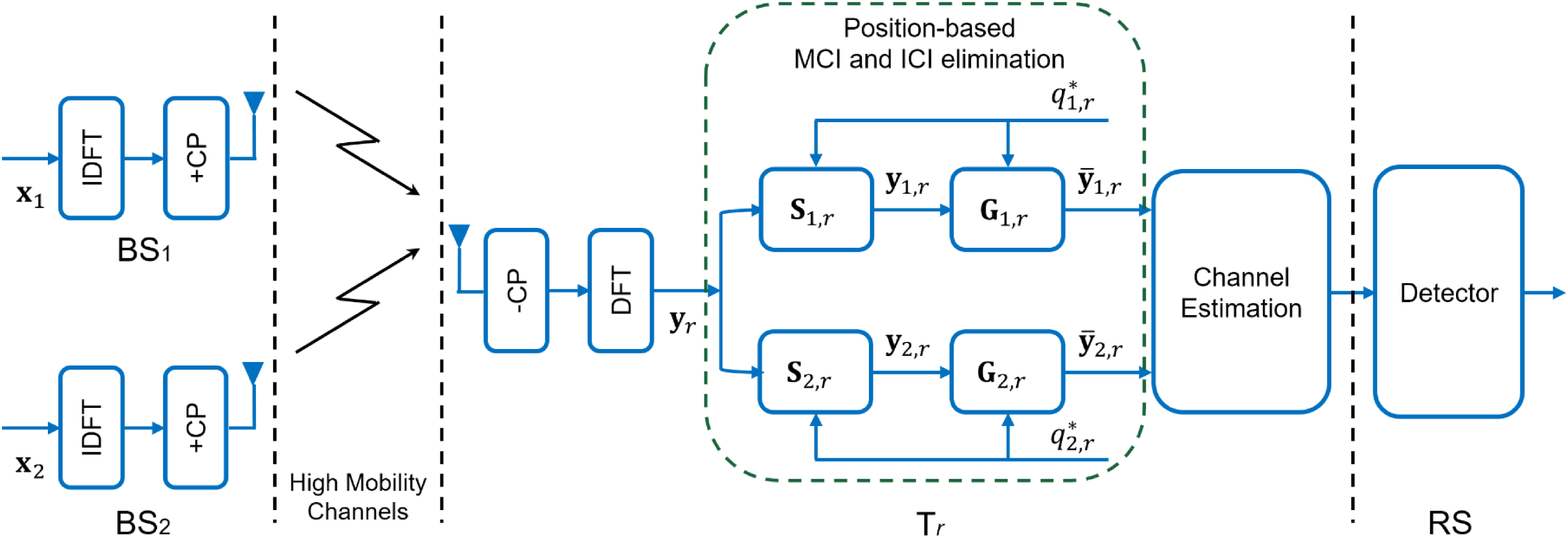}\\
  \caption{The structure of the position-based MCI and ICI elimination for the $r$-th receive antenna.}\label{RF}
\end{figure*}

 We now propose a new position-based MCI and ICI elimination method for each receive antenna, which is shown as \reffig{RF}.
Different to the methods in \cite{Aboutorab12} and \cite{Hijazi10} based on iterative process, which may incur large iteration times for high Doppler shift and suffer from a performance degradation due to the error propagation, the proposed method can eliminate both the MCI and ICI before channel estimation by utilizing the train position information, without any iterative process. In addition, unlike \cite{Ren15} which only considers the CE-BEM \cite{Kannu05}, we consider a general BEM in this work.

\subsubsection{MCI elimination}

In \reffig{RF}, for the $r$-th receive antenna in the overlap $A_2-C_1$, after removing the CP and performing the DFT modulation, the received signal ${\bf y}_r$ in (\ref{eq9}) can be rewritten as
\begin{align}
  {\bf y}_r
  & = \sum^{T}_{t=1} {{\bf D}}{\bf \Delta}_{t,r}{\bf x}_t+{\bf n}_r,
\end{align}
where ${\bf D} = [{\bf D}_0,...,{\bf D}_q,...,{\bf D}_{Q}]\in{\mathbb{C}}^{K\times(Q+1)K}$, and ${\bf \Delta}_{t,r} = [{\bf \Delta}^T_{t,r,0},...,{\bf \Delta}^T_{t,r,q},...,{\bf \Delta}^T_{t,r,Q}]^T\in{{\mathbb C}}^{(Q+1)K\times K}$.

Then, ${\bf y}_r$ is transmitted to the designed position-based MCI elimination module ${\bf S}_{t,r}$, where $t =1,2$.  The main idea of ${\bf S}_{t,r}$ is to split up the signals transmitted from different BSs by utilizing their different Doppler shifts.
The designed ${\bf S}_{t,r}$ is a $K\times (Q+1)K$ zero matrix with a $K\times K$ all-one matrix at the position corresponding to $q^*_{t,r}$, which can be represented as
\begin{equation}
  {\bf S}_{t,r} = [{\bf 0},...,\underbrace{{\bf 1}_K}_{q^*_{t,r}},...,{\bf 0}].
\end{equation}
In specific, we have
\begin{equation}
{\bf S}_{t,r}(:, {\bf s}_{t,r}) = {\bf 1}_K,
\end{equation}
where ${\bf s}_{t,r} = [q^*_{t,r}K,q^*_{t,r}K+1,...,(q^*_{t,r}+1)K-1]$ is a ${1\times K}$ vector collecting the column indices of the all-one matrix in ${\bf S}_{t,r}$, and $q^*_{t,r}\in\{0,1,...,Q\}$.
Note that $q^*_{t,r}$ is related to $f_{t,r}$ and their relationship is given as (\ref{q_fd}).

Denote ${\bf y}_{t,r}$ as the signal transmitted by $BS_t$ and received at the $r$-th receive antenna.
After passing ${\bf S}_{t,r}$, ${\bf y}_{t,r}$ can be obtained  as
\begin{align}
  {\bf y}_{t,r}
  & = {\bf S}_{t,r}\star  \sum^{T}_{t=1} {{\bf D}}{\bf \Delta}_{t,r}{\bf x}_t+{\bf n}_{t,r},\\
  & = {\bf S}_{t,r}\star{\bf D}{\bf \Delta}_{t,r}{\bf x}_t + \sum^T_{\nu=1,\nu\neq t}{\bf S}_{t,r}\star {\bf D}{\bf \Delta}_{\nu,r}{\bf x}_{\nu}  +{\bf n}_{t,r},\label{IAI2}\\
  & = {\bf D}_{{q}^*_{t,r}}{\bf \Delta}_{t,r,{q^*_{t,r}}}{\bf x}_t + \sum^T_{\nu=1,\nu\neq t}\underbrace{{\bf D}_{q^*_{t,r}} {\bf \Delta}_{\nu,r,q^*_{t,r}}}_{{\bf J} = {\bf 0}}{\bf x}_{\nu}  +{\bf n}_{t,r}, \label{IAI1}\\
  & = {\bf D}^*_{t,r}{\bf \Delta}^*_{t,r}{\bf x}_t + {\bf n}_{t,r},\label{IAI}
\end{align}
where $\star$ is the dot product operator, ${\bf n}_{t,r}$ is the equivalent noise vector after passing ${\bf S}_{t,r}$, and $t = 1,2$. Note that we have ${\bf J} = {\bf 0}$ ($\nu \neq t$) in (\ref{IAI1}) since its corresponding BEM coefficients ${\bf c}_{\nu,r,q^*_{t,r}}$  are non-dominant for ${\bf H}_{\nu,r}$, i.e., ${\bf \Delta}_{\nu,r,q^*_{t,r}} = {\text{diag}}\{{\bf F}_L{\bf c}_{\nu,r,q^*_{t,r}}\}  = {\bf 0}$.
Whereas, according to Lemma 1, the dominant coefficients of ${\bf H}_{\nu, r}$ exist in ${\bf c}_{\nu,r,{q}^*_{\nu,r}}$ and thus correspond to  ${\bf \Delta}_{\nu,r,q^*_{\nu,r}} \neq{\bf 0}$.
In (\ref{IAI1}), we also have
\begin{equation}
{\bf S}_{t,r}\star{{\bf D}}{\bf \Delta}_{t,r} = {\bf D}_{{q}^*_{t,r}}{\bf \Delta}_{t,r,{q^*_{t,r}}}.
\end{equation}
This is because that ${\bf S}_{t,r}$ can be seen as a selection matrix which selects the dominant basis matrix
${\bf D}_{{q}^*_{t,r}}$ from $\bf D$ and the dominant coefficient matrix ${\bf \Delta}_{t,r,{q^*_{t,r}}}$ from ${\bf \Delta}_{t,r}$ corresponding to $q^*_{t,r}$.
Since $q^*_{t,r}$ is related to the antenna position $\alpha_{t,r}$ as (\ref{q_a}), ${\bf S}_{t,r}$ is called as position-based MCI eliminator.
From (\ref{IAI}), it is easy to find that the received signal after passing ${\bf S}_{t,r}$ is free of the MCI.
In addition, when the $r$-th receive antenna is out of the overlap, we find that (\ref{IAI}) turns into (\ref{eq13}),
which means that ${\bf S}_{t,r}$ can be seen as a process of selecting the position-based channel model.

\begin{figure*}[t!]
  \centering
  \includegraphics[width=5.6in]{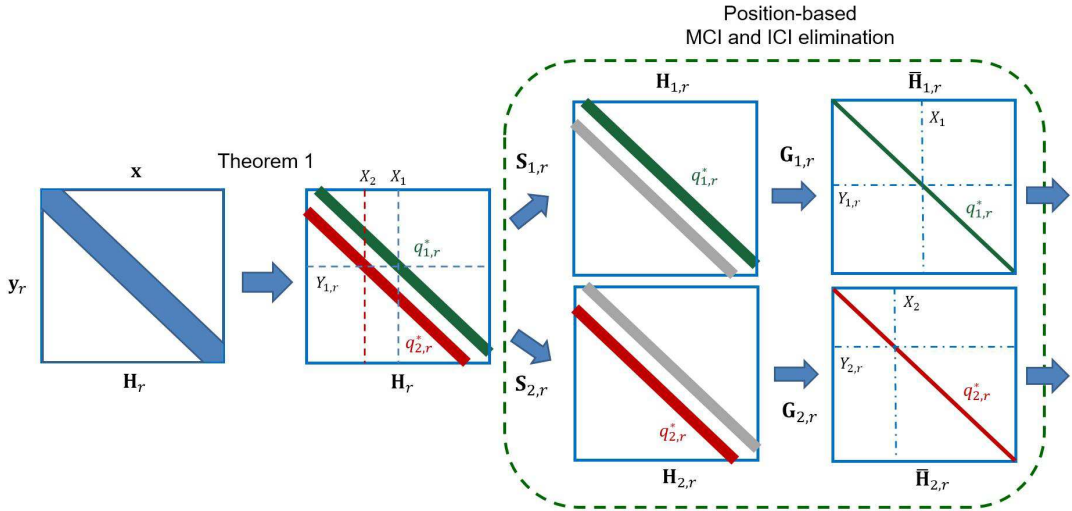}\\
  \caption{The structure of ${\bf H}_r$ with the position-based MCI and ICI elimination method. (The color parts denote the non-zero entries, and the white parts denote the zero entries. The green solid lines denote the entries corresponding to ${\bf D}^*_{1,r}$ with $q^*_{1,r}$, and the red solid lines denote the entries corresponding to ${\bf D}^*_{2,r}$ with $q^*_{2,r}$. )}\label{ici}
\end{figure*}

\subsubsection{ICI elimination}

After passing the proposed MCI eliminator, the MCI-free signial at the $r$-th receive antenna is obtained as (\ref{IAI}). In (\ref{IAI}), since ${\bf D}^*_{t,r} = {\bf D}_{q|q={q}^*_{t,r}}$ is a deterministic square matrix for the given BEM and $q^*_{t,r}$ can be directly calculated for a given antenna position as (\ref{q_a}), we can thus design a matrix ${\bf G}_{t,r}$ to eliminate the ICI incurred by ${\bf D}^*_{t,r}$ at the receiver, i.e., ${\bf G}_{t,r}{\bf D}^*_{t,r} = {\bf I}$.
In addition, as ${\bf D}^*_{t,r} = {\bf F}{\text{diag}}\{{\bf b}_{{q}^*_{t,r}}\}{\bf F}^H$ is nonsingular for most BEMs \cite{Tang07}, ${\bf G}_{t,r}$ can be directly calculated as  ${\bf G}_{t,r} = {{\bf D}^*_{t,r}}^{-1}$.
Note that ${\bf G}_{t,r}$ is also position-based since it corresponds to the location-related index $q^*_{t,r}$.
In practice, ${\bf G}_{t,r}$ can be pre-designed for $\{{\bf D}_q\}^Q_{q=0}$ and pre-stored at the receiver.
For the $r$-th receive antenna at a certain position $\alpha_{t,r}$, the corresponding ${\bf G}_{t,r}$ is selected according to $q^*_{t,r}$.

After passing ${\bf S}_{t,r}$, the obtained MCI-free signal $ {\bf y}_{t,r}$ is transmitted to ${\bf G}_{t,r}$ to eliminate the ICI, which can be represented as
\begin{align}
  \bar{\bf y}_{t,r} & = {\bf G}_{t,r} {\bf y}_{t,r},\\
  & =\underbrace{  {\bf G}_{t,r}{\bf D}^*_{{t,r}}}_{\bf I}{\bf \Delta}^*_{{t,r}}{\bf x}_t + \bar{\bf n}_{t,r},\\
  & = {\bf \Delta}^*_{{t,r}}{\bf x}_t + \bar{\bf n}_{t,r},\label{G}
\end{align}
where $\bar{\bf y}_{t,r}$ is the received signal vector after passing ${\bf G}_{t,r}$, and $\bar{\bf n}_{t,r} ={\bf G}_{t,r}{\bf n}_{t,r}$ is the equivalent noise vector after passing ${\bf G}_{t,r}$. Since ${\bf \Delta}^*_{{t,r}}$ is a diagonal matrix consisting of the dominant channel coefficients of ${\bf H}_{t,r}$, it is easy to find that  (\ref{G}) is free of the ICI.
In this way, we can eliminate both the MCI and the ICI at the receive antenna.
Note that these aforementioned analyses and conclusions are not restricted to any specific BEM.
In addition, the proposed ICI elimination method can be directly applied to the case when the receive antenna is out of the overlap as (\ref{eq13}).

\emph{Remark 1:} Different to the ICI elimination method in \cite{Ren15} which can only get the ICI-free signal  for the CE-BEM, the conclusion that (\ref{G}) is ICI-free holds for a general BEM. This is  because that
\cite{Ren15} utilizes the property that ${\bf D}^*_{t,r}$ is a permutated identity matrix for the CE-BEM (while ${\bf D}^*_{t,r}$ is approximately banded for most BEMs \cite{Tang07}), whereas,  in this work, ${\bf G}_{t,r}$ is designed for a general BEM to get the ICI-free signal.

\subsubsection{An example}

Here we give an example for better clarification. For the $r$-th receive antenna at the overlap, we plot the structure of its frequency domain channel matrix ${\bf H}_{r}$ with the proposed method in \reffig{ici}. 
 In specific, here we consider the GCE-BEM \cite{Leus04}, which is an improved model of CE-BEM with better modeling performance and robustness for high mobility \cite{Tang07}.
In \reffig{ici}, the columns of ${\bf H}_r$ denote the subcarriers of the pilots and data transmitted from BSs,  
and the rows denote the subcarriers of the received signals at the $r$-th receive antenna.
For the $r$-th receive antenna at the overlap, ${\bf H}_r$ is approximately banded, which is shown as the blue parts.
 With Theorem 1, ${\bf H}_r$ can be represented as a matrix with two approximately banded entries corresponding to $q^*_{1,r}$ (the green band) and $q^*_{2,r}$ (the red band), respectively.
 Denote $X_t$ as a signal transmitted by $BS_t$ at a certain subcarrier, and $Y_{t,r}$ as the desired received signal of $X_{t}$, where $t = 1,2$.
It can be observed that  $Y_{1,r}$ (the blue dash line) suffers from both the MCI from $X_2$ (the red dash line) and the ICI caused by the neighbouring subcarriers of the desired $X_{1}$ (the blue dash line).
Then, after passing ${\bf S}_{1,r}$ and ${\bf S}_{2,r}$ respectively, ${\bf H}_r$ is separated  into ${\bf H}_{1,r}$ and ${\bf H}_{2,r}$, where the grey parts denotes the eliminated MCI from the other cell.
Note that ${\bf H}_{1,r}$ and ${\bf H}_{2,r}$ still suffer from the ICI.
Next, 
after passing ${\bf G}_{t,r}$, ${\bf H}_{t,r}$ turns into the green (or red) line corresponding to $q^*_{t,r}$, which is because that ${\bf G}_{t,r}$ is designed for ${\bf D}^*_{t,r}$ and the dominant coefficients in ${\bf \Delta}^*_{t,r}$ alone describe the channel with the Doppler shift $f_{t,r}$.
 It can be observed that the desired signal $X_t$ is free of the ICI and received as $Y_{t,r}$ in $\bar{\bf H}_{t,r}$ (the blue dash dot lines).
In addition, different to the method in \cite{Ren15}, the subcarrier permutation  in $\bar{\bf H}_{t,r}$ caused by the large Doppler shift is also eliminated by ${\bf G}_{t,r}$.
In this way, the MCI and the ICI are eliminated at each receive antenna by the proposed method.

\setcounter{TempEqCnt}{\value{equation}} 
\setcounter{equation}{36}                           
\begin{figure*}[!t]
\begin{align}
\begin{bmatrix}
    \bar{\bf y}_{1,r}({\bf w}_1)\\
    \bar{\bf y}_{2,r}({\bf w}_2)
\end{bmatrix} & = \begin{bmatrix}
    {\bf \Lambda}_1({\bf w}_1,:) & \\
    & {\bf \Lambda}_2({\bf w}_2,:)
\end{bmatrix}
\begin{bmatrix}
    {\bf c}^*_{1,r}\\
    {\bf c}^*_{2,r}
\end{bmatrix}
 +\begin{bmatrix}
   \bar{\bf n}_{1,r}({\bf w}_1)\\
    \bar{\bf n}_{2,r}({\bf w}_2)
\end{bmatrix} ,\\
 & = \underbrace{\begin{bmatrix}
    \text{diag}\{{\bf x}_1({\bf w}_1)\}{\bf F}_L({\bf w}_1,:)& \\
    & \text{diag}\{{\bf x}_2({\bf w}_2)\}{\bf F}_L({\bf w}_2,:)
\end{bmatrix}}_{{\bf \Psi}_r}
\begin{bmatrix}
    {\bf c}^*_{1,r}\\
    {\bf c}^*_{2,r}
\end{bmatrix}
 +\begin{bmatrix}
   \bar{\bf n}_{1,r}({\bf w}_1)\\
    \bar{\bf n}_{2,r}({\bf w}_2)
\end{bmatrix}. \label{CE}
\end{align}
\hrulefill
\end{figure*}
\setcounter{equation}{\value{TempEqCnt}}

\section{Compressed Channel Estimation and Pilot Pattern Design}
In this section, after briefly reviewing some backgrounds of CS,  a low-complexity compressed channel estimation method with the optimal pilot pattern design is developed for the considered multi-cell HST system. 
In addition, the complexity and the summary of the proposed method are given.

\subsection{CS Theory}
CS is a revolutionary technique which can recover sparse signals from the undersampled measurements \cite{Candes06}. Let ${\hat{\bf{x}}}\in \mathbb{C}^M$ be an unknown signal vector and it can be represented as an $S$-sparse vector ${\bf a} \in {\mathbb{C}}^{N}$ with a known matrix ${\bf{\Phi}}\in {\mathbb{C}}^{M \times N}$, i.e.,  ${\hat{\bf{x}}}=\bf{\Phi}\bf{a}$ and  $\|{\bf{a}}\|_{{{{\ell }_{0}}}}=S\ll N$.
With a given measurement matrix ${\bf{\Psi}}\in {\mathbb{C}}^{V \times M}$,
CS aims to recover $\bf{a}$ correctly from the observed vector  ${\hat{\bf y}} \in {\mathbb{C}}^{V}$ as
\begin{equation}
{\hat {\bf{{y}}}}={\bf{\Psi}}{\hat{ \bf{{x}}}} + {\bm{\eta}} =\bf{\Psi}\bf{\Phi}\bf{a} + {\bm{\eta}}, \label{CS}
\end{equation}
where  ${\bm{\eta}} \in  {\mathbb{C}}^{V}$ is a noise vector.
The existing work \cite{Elad07}  indicates that a lower average coherence of (\ref{CS}) leads to a more accurate recovery of $\bf a$. 
The average coherence of a matrix has the following definition \cite{Elad07}.
\begin{definition}
The average coherence of a matrix $\bf{Z}$ is defined as the average of all absolute and  normalized inner products between two arbitrary columns in $\bf{Z}$ that are above $\delta$ ($0<\delta<1$), i.e.,
\begin{equation}
\mu _\delta \{ {\bf{Z}}\}  = \frac{{\sum\limits_{i \ne j} {\left( {\left| {z_{ij} } \right| \ge \delta} \right) \cdot \left| {z_{ij} } \right|} }}{{\sum\limits_{i \ne j} {\left( {\left| {z_{ij} } \right| \ge \delta} \right)} }},
\end{equation}
where  $z_{ij} = \tilde{\bf{ z}}^H_i \tilde{\bf{ z}}_j$, $\tilde {\bf{ z}}_i ={\bf{ z}}_i/\|{\bf{ z}}_i\|_{\ell_2} $, ${\bf{z}}_i$ denotes the  $i$-th column of $\bf{Z}$, and
\begin{equation}
(x\geq y)=\left\{\begin{matrix}
  &1,~&x\geq y ,\\
  &0,~&x<y.
\end{matrix}\right.
\end{equation}
\end{definition}

\subsection{Compressed Channel Estimation with Pilot Design}
In this work,  we consider the case that receive antennas estimate their channels individually.
To estimate the channel coefficients, let us rewrite (\ref{G}) as
\begin{align}
  \bar{\bf y}_{t,r} 
  & = {\bf \Lambda}_t{\bf c}^*_{t,r} +\bar{\bf n}_{t,r}, \label{eq37}
\end{align}
where ${\bf \Lambda}_t = \text{diag}\{{\bf x}_t\}{\bf F}_L$.

 For the $r$-th receive at the overlap, since it receives both the signals transmitted by $BS_1$ and $BS_2$,  a low-complexity channel estimation method is developed to jointly estimate the channels  $\{{\bf H}_{t,r}\}^{T=2}_{t=1}$ with the pilot patterns $\{{\bf w}_{t}\}^{T=2}_{t=1}$,   which can be represented as (\ref{CE}).
Note that since $\bar{\bf y}_{1,r}$ has been separated from $\bar{\bf y}_{2,r}$ by the proposed method as (\ref{G}), we can write the channel estimation problem as (\ref{CE}).
In (\ref{CE}), the term ${\bf \Psi}_{r}$ denotes the measurement matrix of the considered problem, which is only related to the transmitted signal ${\bf x}_t$ and the pilot pattern ${\bf w}_{t}$, where $t = 1,2$.
So far, we transfer the problem of estimating the frequency domain channel matrix $\{{\bf H}_{t,r}\}^{T=2}_{t=1}$ to estimating the position-based dominant coefficients  $\{{\bf c}^*_{t,r}\}^{T=2}_{t=1}$.
The number of channel coefficients that need to be estimated is dramatically reduced from $TK^2$ to $TL$, which highly reduces the estimation complexity.
In addition, when the receive antenna is out of the overlap as (\ref{eq13}), the channel estimation problem can be still represented as (\ref{CE}) with ${\bf c}^*_{\nu,r} = {\bf 0}$ for $\nu \neq t$.

Next, with the conclusion of \cite{Elad07}, the pilot patterns $\{{\bf w}_t\}^{T}_{t=1}$ are optimized to minimize the average coherence of (\ref{CE}).
In this work, we assume that the pilot symbols have the same constant amplitude at each BS, i.e.,
\addtocounter{equation}{2}  
\begin{equation}
|X_t(w_{t,p})|^2 = A, ~~\forall~{w}_{t,p}\in {\bf w}_t,
\end{equation}
where $t = 1,2$ and $p=1,2,...P$.
Then, the optimization problem is formulated as
\begin{align}
\left\{{\bf w}^*_1, {\bf w}^*_2\right\} &= \arg\min_{{\bf w}_1,{\bf w}_2} \mu_\delta\{{\bf \Psi}_r\}, \\
& = \arg\min_{{\bf w}_1,{\bf w}_2} \mu_\delta\left\{\begin{bmatrix}
    {\bf \Lambda}_1({\bf w}_1,:)& \\
    & {\bf \Lambda}_2({\bf w}_2,:)
\end{bmatrix}\right\},\label{eqoj}
\end{align}
where ${\bf w}^*_t$ represents the optimal pilot pattern for $BS_t$, and $t = 1,2$.
In (\ref{eqoj}), according to Definition 2, it is easy to find that minimizing  $\mu_\delta\{{\bf \Psi}_r\}$
equals to individually minimizing $\mu_\delta\{ {\bf \Lambda}_1({\bf w}_1,:)\}$   and $ \mu_\delta\{ {\bf \Lambda}_2({\bf w}_2,:)\}$. Thus, (\ref{eqoj}) can be rewritten as
\begin{align}
{\bf w}^*_t  &= \arg\min_{{\bf w}_t} \mu_\delta\{{\bf \Lambda}_t({\bf w}_t,:)\}, \label{eq42}\\
& =   \arg\min_{{\bf w}_t} \mu_\delta\{ \text{diag}\{{\bf x}_t({\bf w}_t)\}{\bf F}_L({\bf w}_t,:)\},
\end{align}
where $t = 1, 2$.
From Definition 2, since the average coherence is independent of the constant amplitude, the objective function can be further expressed as
\begin{align}
\mu_\delta\{ \text{diag}\{{\bf x}_t({\bf w}_t)\}{\bf F}_L({\bf w}_t,:)\}
&=  \mu_\delta\left\{A {{\bf{F}}_L}({{\bf{w}}_t},:)\right\},\\
&=  \mu_\delta\left\{ {{\bf{F}}_L}({{\bf{w}}_t},:)\right\}.\label{eq44}
\end{align}

Therefore, the optimization problem in (\ref{eqoj}) is converted to the following problem as
\begin{align}
{\bf w}^*_t   =   \arg\min_{{\bf w}_t} \mu_\delta\left\{ \text{diag}\{{\bf F}_L({\bf w}_t,:)\right\}, ~~~&t = 1,2.\label{eqoj2}
\end{align}
From (\ref{eqoj2}), we find that ${\bf w}^*_t $ is independent of the train position and speed, the number of BSs and receive antennas, and even the fast variation of the Doppler shift $f_{t,r}$,
which means that ${\bf w}^* = \{{\bf w}_t^*\}^T_{t=1} $  is global optimal for the considered multi-cell multi-antenna HST system.

\subsection{Pilot Pattern Design Algorithm}

\begin{algorithm}[!t]\small
 \caption{{\bf{:}} Pilot Pattern Design Algorithm}
  \begin{algorithmic}[1]
\Require Initial pilot pattern ${\bf w} = {\bf w}_t$;
\Ensure Optimal pilot pattern ${\bf w}^*  = \hat{\bf w}^{(MP)}$;\\
{\bf Procedure:}
\raggedright
\State {\bf Initialization}: Set $Iter = M\times P$, set ${\bf \Gamma}={\bf 0}$ and $\Gamma[0,0]=1$, set $\kappa=0$ and $\iota = 0$.\\
\State {\bf For}~$ n = 0, 1,..., M-1 $\\
\State {\bf For}~$ \bar{p}= 0, 1,..., P-1 $\\
\State $m= n \times P+ \bar{p}$;
\\
        \State generate $\tilde{\bf{w}}^{(m)}$ with operator ${\bf{w}}^{(m)}\Rightarrow\tilde{\bf{w}}^{(m)}$;
        \If {$\mu_\delta\{ {{\bf{F}}_L}({\tilde{\bf{w}}^{(m)}},:)\} < \mu_\delta\{{{\bf{F}}_L}({{\bf{w}}^{(m)}},:)\}$}
         \State ${\bf{w}}^{(m+1)} = {{\tilde{\bf{w}}}^{(m)}}$;~  $\kappa = m+1$;
        \Else
         \State ${{{\bf{w}}}^{(m+1)}} = {{{\bf{w}}}^{(m)}}$;
        \EndIf
        \\
        \State ${\bf{\Gamma}}[m+1] = {\bf{\Gamma}}[m] + \eta[m]({\bf{U}}[m+1] - {\bf{\Gamma}}[m])$, with  $\eta[m]=\frac{1}{m+1}$;
        \If {${{\Gamma}}[m+1,\kappa] > {{\Gamma}}[m+1,\iota]$}
            \State $\hat{\bf{w}}^{(m+1)} = {\bf{w}}^{({m+1})}$; ~$\iota \Leftarrow \kappa$;
        \Else
            \State $\hat{\bf{w}}^{({m+1})} = \hat{\bf{w}}^{({m})}$;
        \EndIf
\State {\bf End For}~($\bar{p}$)
\State {\bf End For}~($n$)
\end{algorithmic}
\end{algorithm}

With the proposed interference elimination and channel estimation method, we turn the pilot design problem into  (\ref{eqoj2}). This problem can be directly solved by the pilot design algorithm proposed in our previous work \cite{Ren15}, which is briefly presented as Algorithm 1.

In Algorithm 1, $Iter = M \times P$ denotes the total iteration times and $M$ denotes the number of pilot patterns. Define ${\bf{w}}^{(m)}$, $\tilde{\bf{w}}^{(m)}$, and $\hat{\bf{w}}^{(m)}$  as some pilot patterns at the $m$-th iteration.
The vector ${\bf{\Gamma}}[m] = [\Gamma[m,1],\Gamma[m,2],...,\Gamma[m,MP]]^T$ presents the state occupation probabilities with elements ${{\Gamma}}[m,\kappa] \in [0,1]$ and $\sum_{\kappa}{{\Gamma}}[m,\kappa] =1$.
${\bf{U}}[m]$ is an ${MP\times 1}$ vector with the $m$-th element as 1 and other elements as zero.
The operator ${\bf{w}}^{(m)}\Rightarrow\tilde{\bf{w}}^{(m)} $ means that replacing the $\bar{p}$-th element of ${\bf{w}}^{(m)}$ as a random element which is not included in ${\bf{w}}^{(m)}$ at the $m$-th iteration.
For each iteration,  a candidate with a smaller average coherence is allocated for the next iteration and 
${\bf{\Gamma}}[m+1]$ is updated with the decreasing step size $\eta[m] = 1/(m+1)$.
The pilot pattern with the largest element in  ${\bf{\Gamma}}[m+1]$ is updated as the current optimal pattern.
The convergence of Algorithm 1 is given in \cite{Ren15}.

\emph{Remark 2:}
Different to the pilot design algorithm in \cite{Ren15},  in this work, we do not need to design a specific receive pilot pattern for each receive antenna at different positions, which will certainly reduce the complexity. In \cite{Ren15}, each receive antenna needs to design its own receive pilot pattern according to its instant position to eliminate the subcarrier permutation caused by the Doppler shift.
In this work,  the subcarrier permutation is eliminated by the designed ${\bf G}_{t,r}$.
 In addition, for the considered multi-cell system, \cite{Ren15} needs additional $P$ guard pilots to eliminate the MCI. Whereas, in this work, the MCI is eliminated by ${\bf S}_{t,r}$ without the help of guard pilots, resulting in high spectrum efficiency.

\subsection{Complexity Analysis} \label{Complex}
The complexity of the proposed method is discussed in the term of the needed multiplications, which mainly depends on the interference elimination and the pilot pattern design.
\begin{itemize}
  \item
      For the MCI elimination, with the designed ${\bf S}_{t,r}$, the $r$-th antenna obtains the signal from each BS with (\ref{IAI}). This process requires $T(Q+1)K^2$ complex multiplications.
      Note that ${\bf S}_{t,r}$ for any given position can be off-line pre-designed  for $q \in \{0,1,...,Q\}$ and selected according to $q^*_{t,r}$.
      For the ICI elimination, since ${\bf G}_{t,r}$ can be also off-line pre-designed for $\{{\bf D}_{q}\}^Q_{q=0}$ of the considered BEM, this process requires $TK^2$ complex multiplications. Therefore, the proposed interference elimination method requires $T(Q+2)K^2$ complex multiplications in total.
  \item 
      For the pilot pattern design, Algorithm 1 is an off-line operation with given system parameters. In addition, comparing with the pilot design algorithm in \cite{Ren15}, the proposed method does not need to design the receive pilot pattern for each receive antenna at different positions, which further reduces the system complexity.
\end{itemize}

Therefore, the proposed method needs $T(Q+2)K^2$ complex multiplications in total. In practice, since the constant system parameters $T$ and $Q$ are much smaller than $K$, the complexity of the proposed method is ${\mathcal O}(K^2)$.
In contrast, the complexity of the methods in \cite{Campos13} and \cite{Hijazi10} are  ${\mathcal O}(K^2)$
and ${\mathcal O}(K^3)$, respectively, which,  however, only consider the ICI and will need additional complexity to combat the MCI.  In addition, as the GPS has been widely equipped in current HST systems \cite{Pascoe09}, it is convenient to obtain the train's position and speed information.
This makes the proposed method feasible for implementation in practical systems without greatly increasing the system complexity.


\subsection{Scheme Summary and Comparison}
 Now we make a briefly summary of the proposed scheme in the considered multi-cell HST system, which is given as follows:
 \begin{enumerate}
 	 \item With given system parameters, we first utilize Algorithm~1 to obtain  ${\bf w}^*$ and store it at the BSs and the receive antennas.  Then, for the considered BEM channel model, both ${\bf S}_{t,r}$ and ${\bf G}_{t,r}$ are off-line pre-designed for $q = 0,1,...,Q$ and stored at the receive antennas.
   \item During the system run, the received signals on the train are sent to the proposed MCI and ICI eliminator, where ${\bf S}_{t,r}$ and ${\bf G}_{t,r}$ are selected for each receive antenna according to $q^*_{t,r}$. The dominant index ${q^*_{t,r}}$ is calculated from  (\ref{q_fd}) or (\ref{q_a}) with the help of the GPS.
       The interference eliminated signal at each receive antenna is given as (\ref{G}).
   \item Finally, each receive antennas utilizes ${\bf w}^*$ to jointly estimate the high mobility channels of different BSs as (\ref{CE}). The estimated channels are then sent to the RS for further operation.
 \end{enumerate}

Next, we compare the proposed method and the method in \cite{Ren15}.
Consider a multi-cell HST system,  the receive antennas receive the signals from different cells at the overlap of two adjacent cells, incurring the MCI.
To combat the MCI,  the total $P$ pilots in \cite{Ren15} can be divided into two subsets and each BS uses one subset to transmit the optimal pilot to the receive antenna for channel estimation.
These  subsets are orthogonal in the frequency domain, i.e., one subset put the guard pilots at the subcarriers of the other subset.
Thus, each BS only has $P/2$ effective pilots, which highly reduces the spectrum efficiency.
Whereas, for the proposed method, each BS has $P$ effective pilots since the MCI is eliminated by selecting different Doppler shifts without needing additional guard pilots.
Furthermore, the ICI elimination scheme in \cite{Ren15} can only get the ICI-free pilots in the CE-BEM.
For other BEMs, e.g., the GCE-BEM, the receive pilots   in \cite{Ren15} will still suffer from the ICI since the ${\bf D}^*_{t,r}$ is no more strictly  banded, which incurs residual ICI and needs more complexity to further mitigate the ICI.
In contrast, in this work, since ${\bf G}_{t,r}$ is pre-designed for the considered BEM, the ICI-free pilots can be obtained at the receivers.
In addition, to eliminate the subcarrier permutation incurred by the large Doppler shift, the method in \cite{Ren15} needs additional complexity to design the receive pilot pattern for each receive antenna according to its instant position. Whereas,  the subcarrier permutation is also eliminated by the pre-designed ${\bf G}_{t,r}$ in this work.

Here we discuss the impacts of the GPS location error on the proposed method.
Note that the designed pilot pattern is independent of the train position.
For the proposed MCI and ICI elimination method, since ${\bf S}_{t,r}$ and ${\bf G}_{t,r}$ are only related to $q^*_{t,r}$, its performance only depends on the accuracy of $q^*_{t,r}$,  where $q^*_{t,r}$ can be directly calculated by (\ref{q_a}) for a certain $\alpha_{t,r}$.
In  (\ref{q_a}), since $D_0$ and $D_{min}$ are much larger than the location error in practice, it can be found that the location error will not greatly reduce the accuracy of $q^*_{t,r}$.
Moreover, as $q^*_{t,r}$ is a quantized index, it is easy to find that many different train positions may correspond to a same $q^*_{t,r}$. In this way, one may still obtain the correct $q^*_{t,r}$ while there exists a location error, which guarantees the robustness of the proposed method against the location error.
In practice, the location error may has a larger impact on the proposed method when the train is passing the BS, especially at the closest  position, i.e., $B_t$.
When the receive antenna just moves at $B_t$, a larger location error may result in a wrong $q^*_{t,r}$ and the system will select the wrong ${\bf S}_{t,r}$ and ${\bf G}_{t,r}$ for interference elimination, reducing the performance of the proposed method.
However, as the HST always runs at a speed more than 500km/h, the train will pass the BS very quickly which costs a very short time comparing to the whole system runs. In addition, in practical systems, many other positioning systems, e.g., track circuits \cite{Pascoe09}, can be combined to further reduce the location error, which may improve the robustness of the proposed method.  On the other hand, the Doppler shift estimation method in \cite{Hou16} can be also used to support the MCI elimination scheme with location errors.


\section{Simulation Results}

In this section, simulation results are presented to demonstrate the benefits of the proposed method in the considered multi-cell multi-antenna HST system.
The mean square error (MSE) and the bit error rate (BER) are illustrated versus the signal to noise ratio (SNR) and  HST positions.  Here the MSE is given as
\begin{equation}
  \text{MSE} = \frac{1}{\mathcal{I} K^2}\sum^{\mathcal{I}}_{\mathfrak{i}=1}\|{\bf H}^{\mathfrak{i}}_{t,r}-\hat{{\bf H}}^{\mathfrak{i}}_{t,r}\|^2_{F},
\end{equation}
where $\mathcal{I}$ is the number of channel realizations, $\hat{{\bf H}}^{\mathfrak{i}}_{t,r}$ is the estimated channel matrix of ${\bf H}^{\mathfrak{i}}_{t,r}$ at the $\mathfrak{i}$-th realization, and $\|\cdot\|_F$ denotes the Frobenius matrix norm.
Two typical compressed channel estimators are considered:  the basis pursuit (BP) \cite{Chen01} and the orthogonal matching pursuit (OMP) \cite{Pati93}.
The HST system parameters are given in Table~\ref{tb1}, where two receive antennas ($R=2$) are employed at the front and at the rear of the train, respectively.
The train speed is set to 500km/h corresponding to a maximum Doppler shift as $f_{{max}} = 1.088$KHz.
We consider an OFDM system with 512 subcarriers,  30 pilots, the $4$-QAM modulation, the $5$MHz bandwidth, and the $T_d = 1.2$ms packet duration.
In this work, the GCE-BEM is adopted to model the high mobility channels due to its good modeling performance at high Doppler frequencies \cite{Tang07}. The CE-BEM is also considered for comparison.
Each high mobility channel is considered with $L = 64$ taps and $8$ dominant taps.
In addition, the conventional methods presented  in \cite{Ren15}  and  \cite{Cheng13} are included for comparison.

\begin{table}[!t]
\centering
\caption{\label{table1} HST COMMUNICATION SYSTEM PARAMETERS} \label{tb1}
\renewcommand\arraystretch{1.1}
\begin{tabular}{c  c  c} \toprule
{Parameters} & {Variables}&{Values} \\  \toprule
BS cover range  & $R_{BS}$     &    $ 1200$~m      \\
Distance between BSs  & $D_{s}$     &    $ 2000$~m      \\
Max distance of BS to railway  & $D_{max}$    &     $1200$~m \\
Min distance of BS to railway  & $D_{min}$    &     $50$~m     \\

Overlap range     & $D_{c}$     &    $400$~m      \\
HST length & $L_{hst}$ & $240$~m\\

Carrier frequency & $f_c$ & $2.35$~GHz \\
Train speed & $v$    &   $500$~km/h       \\
\bottomrule
\end{tabular}
\end{table}

\subsection{MSE Performances}

\begin{figure}[!t]
\centering
\includegraphics[width=3.4in]{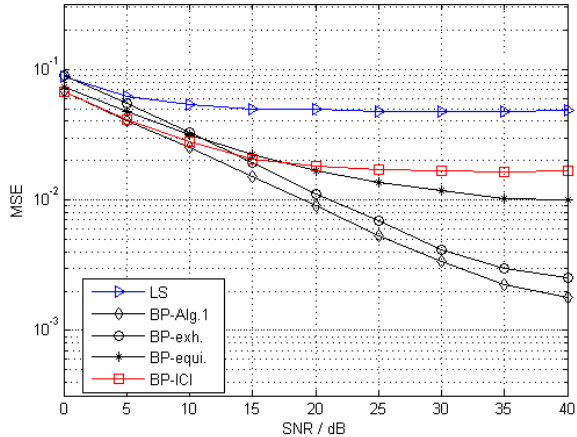}
\caption{MSE performances of  LS and BP estimators in the GCE-BEM at the position $A_1$.}\label{s1}
\end{figure}

\reffig{s1} indicates the MSE performances of different channel estimators for the receive antenna at the position $A_1$  in the GCE-BEM, where the receive antenna suffers a Doppler shift from $BS_1$ as $1.087$KHz.
In this figure, we consider the proposed method with the BP estimator and three different pilot pattern design methods: the pilot pattern obtained by Algorithm 1 (``Alg.1''), the exhaustive pilot design method (``exh.'') in \cite{He10},
and the equidistant pilot pattern (``equi.'')  in \cite{Ma03}.
In addition, the least square (LS) estimator (``LS'') with the pilot pattern in \cite{Ma03} and the proposed interference elimination method is included.
The ICI elimination method in \cite{Ren15} (``BP-ICI'') is also considered with the BP estimator and the pilot pattern designed by Algorithm~1.
All these methods are considered with 30 pilots.
It can be found that the proposed method (``BP-Alg.1'')  can effectively reduce the ICI and achieves better performance than the method in \cite{Ren15} (``BP-ICI'') for the GCE-BEM. 
This is reasonable because that \cite{Ren15} is designed for the CE-BEM and still suffers from the ICI caused by the approximately banded ${\bf D}_q$ for the GCE-BEM, resulting in a performance degradation.

\begin{figure}[!t]
\centering
\includegraphics[width=3.4in]{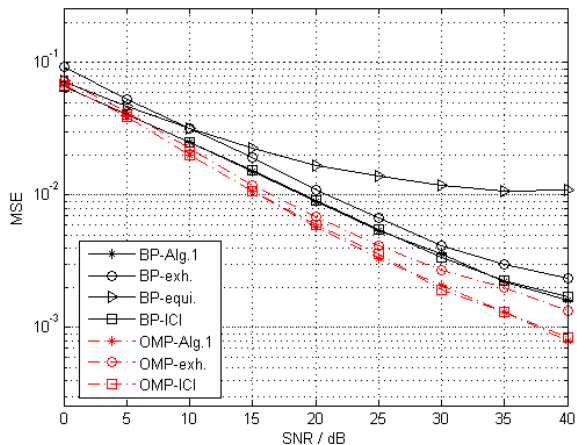}
\caption{MSE performances of  BP and OMP estimators in the CE-BEM at the position $A_1$.}\label{s2}
\end{figure}

\begin{figure}[!t]
\centering
\includegraphics[width=3.4in]{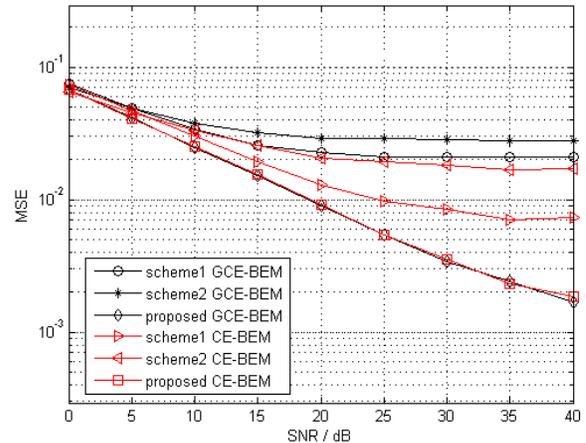}
\caption{Comparisons of the MSE performances of different schemes at the overlap $A_2-C_1$.}\label{s3}
\end{figure}

\reffig{s2} shows the MSE comparison of different estimators for the receive antenna at $A_1$ in the CE-BEM.
As can be seen, the BP and OMP estimators with the proposed interference elimination method (``BP-Alg.1'' and ``OMP-Alg.1") achieve the similar performances as the estimators with the method in \cite{Ren15} (``BP-ICI'' and ``OMP-ICI''), which means that the proposed method is also effective for the CE-BEM.
Note that it has been proven in \cite{Ren15} that its method can get the ICI-free pilots for the CE-BEM.

In \reffig{s3}, we compare the MSE performances of different schemes for the receive antenna at the overlap  $A_2-C_1$, where the receive antenna suffers the Doppler shifts from $BS_1$ as $f_{1,r} = -1.087$KHz and from $BS_2$ as $f_{2,r} = 1.087$KHz, respectively.
Scheme 1 denotes the method in \cite{Ren15} with 48 pilots, where the pilots are divided into two orthogonal subsets to eliminate the MCI, i.e.,  each BS has 24 effective pilots.
Scheme 2 denotes the method in \cite{Cheng13} with 200 pilots ($20$ effective pilots for each BS), where $160$ guard pilots are required for ICI elimination and $20$ guard pilots are needed for MCI elimination.
All the compared schemes are considered with the BP estimator.
In this figure, we consider both the GCE-BEM and the CE-BEM.
For better illustrating the effectiveness of the proposed scheme for different BEMs, here we ignore the modeling error and assume that the channels have the same sparsity in the GCE-BEM and the CE-BEM.
From \reffig{s3},  we find that the proposed scheme with 30 pilots achieves the best estimation performance for both the GCE-BEM and the CE-BEM.
This is because that the proposed method can eliminate the MCI and the ICI without needing any guard pilot, and thus each BS has 30 effective pilots.
Comparing the GCE-BEM and the CE-BEM,  it can be seen that both the performances of the scheme~1 and the scheme~2 are degraded. This is mainly because that these schemes are designed for the CE-BEM and suffer from the residual ICI in the GCE-BEM.
However, it can be noticed that the performances of the proposed scheme for the CE-BEM and the GCE-BEM are almost superimposed, which verifies its robustness  to different BEMs.

\subsection{MSE Performances versus Position}

\begin{figure}[!t]
\centering
\includegraphics[width=3.4in]{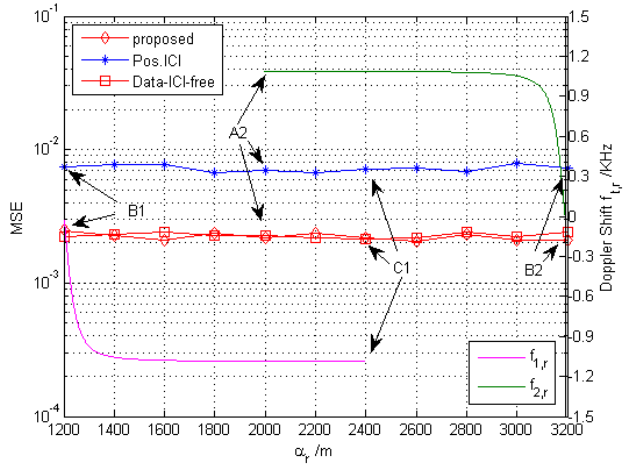}
\caption{MSE performances of BP estimators and the
Doppler shifts versus the antenna position $\alpha_r$ in the CE-BEM, SNR=35dB. $\alpha_r\in[1200,3200]$m for the $r$-th receive antenna moving from $B_1$ to $B_2$.}\label{s4}
\end{figure}

\reffig{s4} illustrates the MSE performances and the Doppler shifts at different receive antenna positions in the CE-BEM, where $f_{1,r}$ is the Doppler shift of $BS_1$, $f_{2,r}$ is the Doppler shift of $BS_2$, and SNR = $35$dB.
For better illustration, here we denote $\alpha_r$ as the distance between the $r$-th receive antenna and $A_1$, and we have $\alpha_r\in[1200,3200]$m for the antenna moving from $B_1$ to $B_2$.
It can be found  that Doppler shifts changes rapidly near $B_1$ and $B_2$, and the receive antenna suffers from two large Doppler shifts at the overlap $A_2-C_1$.
\reffig{s4} includes the proposed method with 30 pilots, the ICI elimination method in \cite{Ren15} with 48 pilots (``Pos.ICI''), i.e., 24 effective pilots for each BS,  and the estimation method where pilots are free of the ICI coming from data subcarriers (``ICI-free''), i.e., data are set as zero.
All these methods are equipped with the BP estimators and Algorithm 1.
As can be seen, the proposed method outperforms the ``Pos. ICI'' method at all positions due to having more effective pilots, and achieves the similar performance as the ``ICI-free'' method since it can effectively eliminate the MCI and the ICI.
In addition,  we find that the proposed method has stable performances while the Doppler shifts significantly change versus the antenna position.
This is mainly because that the proposed MCI and ICI elimination methods and the optimal pilot pattern are all independent of the high mobility and the fast variation of the Doppler shift.

\begin{figure}[!t]
\centering
\includegraphics[width=3.4in]{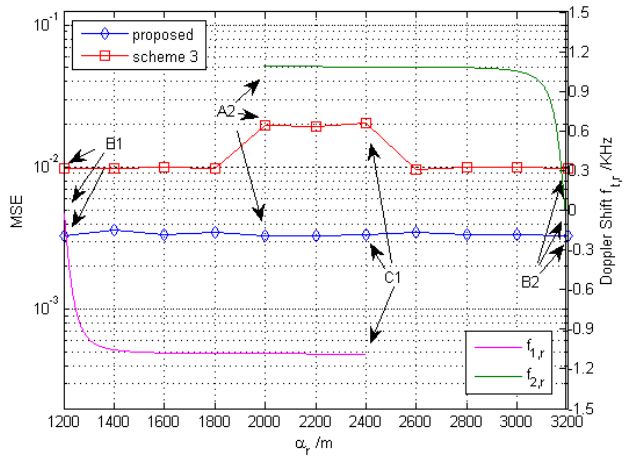}
\caption{MSE performances of different schemes and the
Doppler shifts versus the antenna position  $\alpha_r$ in the GCE-BEM, SNR=25dB. $\alpha_r\in[1200,3200]$m for the $r$-th receive antenna moving from $B_1$ to $B_2$.}\label{s5}
\end{figure}

\reffig{s5} presents the MSE performances of OMP estimators versus the receive antenna position in the GCE-BEM at SNR = $25$dB. As a reference, the Doppler shifts versus the antenna position from $B_1$ to $B_2$ are also plotted.
Scheme 3 is a modified version of the method in \cite{Ren15} with 48 pilots:
when the receive antenna is out of the overlap, i.e., $B_1-A_2$ and $C_1-B_2$, all pilots are used to estimate the channels for each receive antenna; when the receive antenna moves into the overlap $A_2 - C_1$,  the pilots are divided into two orthogonal subsets to eliminate the MCI, i.e., 24 effective pilots for each BS.
As can be seen, the performance of the scheme 3 is degraded for $A_2-C_1$ since some pilots are utilized as the guard pilot to eliminate the MCI.
For $B_1-A_2$ and $C_1-B_2$, it can be noted that the proposed scheme with 30 pilots outperforms the scheme~3 with 48 pilots. This is because that the method in \cite{Ren15} can only get the ICI-free pilot for the CE-BEM and suffers from the residual ICI for the GCE-BEM, which degrades the system performance.
However, it can be observed that the proposed method is  robust to the multi-cell HST system.

\subsection{MSE Performances versus Velocity}

\begin{figure}[!t]
\centering
\includegraphics[width=3.4in]{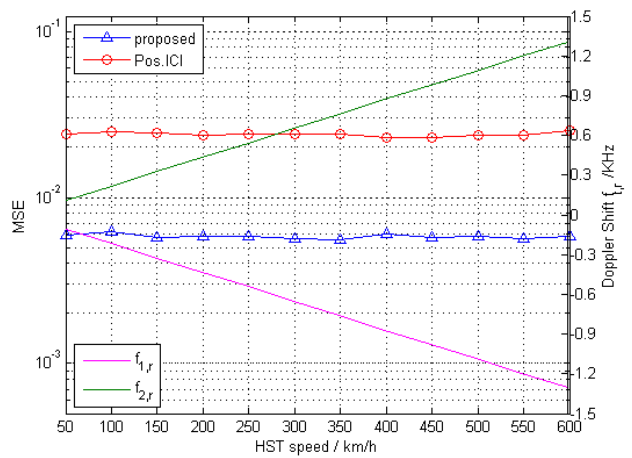}
\caption{MSE performances of OMP estimators and the Doppler shifts versus the train velocity in the GCE-BEM, SNR=20dB. $\alpha_r=2200$m for the $r$-th receive antenna moves to the middle position of the overlap $A_2-C_1$.}\label{fig11}
\end{figure}

\reffig{fig11} compares the MSE performances of the proposed method and the method in \cite{Ren15} (``Pos.ICI'') versus the train velocity, where the $r$-th receive antenna moves to the middle position of the overlap $A_2-C_1$ with $\alpha_r=2200$m.
Both methods  are considered with Algorithm~1 and OMP estimator.
Besides, we also plot the Doppler shifts at the receive antenna, where $f_{1,r}$ and $f_{2,r}$ denote the Doppler shifts caused by $BS_1$ and $ BS_2$, respectively.
As can be seen, both the proposed method and the method in \cite{Ren15} achieve stable performances with increasing train speed, since their interference elimination methods and pilots are independent of the variation of the train speed.
However, it can be found that there exists a performance gap between the proposed method and the conventional one.
This is because that the pilots of \cite{Ren15} are divided into two orthogonal subsets to eliminate the MCI and they also suffer from the residual ICI in the GCE-BEM, resulting in a performance degradation.
Whereas, for different train speeds, the proposed method can directly calculate $q^*_{t,r}$ by (\ref{q_a}) with the help of the GPS, and selects the corresponding ${\bf S}_{t,r}$ and ${\bf G}_{t,r}$ to get the MCI and ICI eliminated signal as (\ref{G}), without needing any guard pilot.
Note that we always have $q^*_{1,r}\neq q^*_{2,r}$ for $f_{1,r}$ and $f_{2,r}$ are with different directions, i.e., $f_{1,r}<0$ and $ f_{2,r}>0$ in \reffig{fig11}.
In addition, as ${\bf S}_{t,r}$ and ${\bf G}_{t,r}$ are only related to  $q^*_{t,r}$, the main conditions for the effectiveness of the proposed method are that there exists a strong LOS propagation path and the receive antenna can obtain the train's instant position and speed information.

\subsection{BER Performances}

\begin{figure}[!t]
\centering
\includegraphics[width=3.4in]{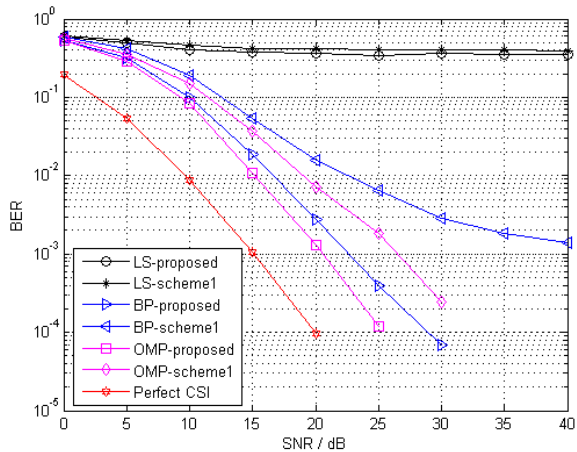}
\caption{BER performances of different channel estimators in the CE-BEM. }\label{s6}
\end{figure}

\begin{figure}[!t]
\centering
\includegraphics[width=3.4in]{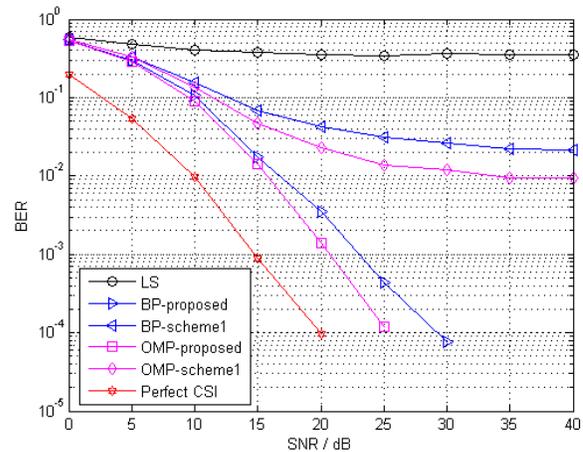}
\caption{BER performances of different channel estimators in the GCE-BEM. }\label{s7}
\end{figure}

\reffig{s6} and \reffig{s7} show the BER performances of the considered system for the CE-BEM and the GCE-BEM, respectively, where the HST moves into the overlap $A_2 - C_1$ and both the two receive antennas receive the signals transmitted from $BS_1$ and $BS_2$.
The proposed method with 30 pilots and the scheme 1 in \cite{Ren15} with 48 pilots are considered with the LS estimator, the BP estimator, and the OMP estimator, respectively.
For all included methods, we consider the zero-forcing (ZF) equalizer. 
In addition, the BER performance with perfect knowledge channel state information (CSI) is also added.
As observed, the proposed method significantly outperforms the scheme 1 for both the BP and the OMP.
For the CE-BEM in \reffig{s6}, we find that the proposed  method outperforms the scheme in \cite{Ren15} for effectively reducing the MCI and improving the effective pilot numbers.
In \reffig{s7}, it can be found that the proposed method still gets better BER performances while the performances of the conventional method are degraded due to the residual ICI in the GCE-BEM.
This is because that the proposed method is not restricted to any specific BEM, whereas, the method in \cite{Ren15} is only designed for the CE-BEM.
Therefore, the proposed method can be well applied to the multi-cell HST system.

\section{Conclusions}

In this paper, we consider the channel estimation and the interference elimination for a mulit-antenna HST communication system in the multi-cell architecture. 
By exploiting the train position information, we show that both the MCI and the ICI can be eliminated at the receive antenna for a general BEM.
Furthermore, we propose a low-complexity compressed channel estimation method with the optimal pilot pattern design for the multi-cell HST system.
The proposed MCI and ICI elimination method and pilot pattern are robust to the high mobility multi-cell OFDM system.
Simulation results verify the effectiveness and the robustness of the proposed method.


\begin{thebibliography}{}

\end{thebibliography}


\begin{thebibliography}{50}


\bibliographystyle{unsrt}

\bibitem{Liu10}
L. Liu, C. Tao, J. Qiu, H. Chen, L. Yu, W. Dong, and Y. Yuan, ``Position-based modeling for wireless channel on high-speed railway under a viaduct at 2.35 GHz," \emph{IEEE Journal on Selected Areas in Communications}, vol. 30, no. 4, pp. 834-845, May 2012.

\bibitem{Wang12}
J. Wang, H. Zhu, and N. J. Gomes, ``Distributed antenna systems for mobile communications in high speed trains,"
\emph{IEEE Journal on Selected Areas in Communications}, vol. 30, no. 4, pp. 675-683, May 2012.


\bibitem{Karimi12}
O. Karimi, J. Liu, and C. Wang, ``Seamless wireless connectivity for multimedia services in high speed trains,'' \emph{IEEE Journal on Selectied Areas in Communnications,} vol. 30, no. 4, pp. 729-739, May 2012.


\bibitem{Wu1}
Q. Wu, W. Chen, M. Tao, J. Li, H. Tang, and J. Wu, ``Resource allocation for joint transmitter and receiver energy efficiency maximization in downlink OFDMA systems,'' \emph{IEEE Transactions on Communications,} vol. 63, no. 2, pp. 416-430, Feb. 2015.

\bibitem{Wu3}
Q. Wu, M. Tao, and W. Chen, ``Joint Tx/Rx Energy-Efficient Scheduling in Multi-Radio Wireless Networks: A Divide-and-Conquer Approach,'' \emph{IEEE Transactions on Wireless Communications,} vol. 15, no. 4, pp. 2727-2740, Apr. 2016.


\bibitem{Mostofi05}
Y. Mostofi and D. Cox, ``ICI mitigation for pilot-aided OFDM mobile systems,'' \emph{IEEE Trans. Wireless Commun.,} vol. 4, no. 2, pp. 765-774, Mar. 2005.

\bibitem{Kwak10}
K. Kwak, S. Lee, H. Min, S. Choi, and D. Hong, ``New OFDM channel
estimation with dual-ICI cancellation in highly mobile channel,'' \emph{IEEE Transactions on Wireless Communications,} vol. 9, no. 10, pp. 3155-3165, Oct. 2010.

\bibitem{Tang07}
Z. Tang, R. Cannizzaro, G. Leus, and P. Banelli, ``Pilot-assisted timevarying channel estimation for OFDM systems,'' \emph{IEEE Transactions on Signal Processing,} vol. 55, no. 5, pp. 2226-2238, May 2007.

\bibitem{Hijazi09}
 H. Hijazi and L. Ros, ``Polynomial estimation of time-varying multipath gains with intercarrier interference mitigation in OFDM systems,'' \emph{IEEE Trans. Veh. Technol.}, vol. 58, no. 1, pp. 140-151, Jan. 2009.


\bibitem{Ma03}
X. Ma, G. Giannakis, and S. Ohno, ``Optimal training for block transmissions over doubly-selective wireless fading channels," \emph{IEEE Trans. Signal Process,} vol. 51, no. 5, pp. 1351-1366, May 2003.


\bibitem{Bajwa10}
W. Bajwa, J. Haupt, A. Sayeed, and R. Nowak, ``Compressed channel sensing: a new approach to estimating sparse multipath channels," \emph{Proceedings of the IEEE,} vol. 98, no. 6, pp. 1058-1076, Jun. 2010.

\bibitem{He10}
 X. He and R. Song, ``Pilot pattern optimization for compressed sensing based sparse channel estimation in OFDM systems,'' in \emph{International Conference on Wireless Communications and Signal Processing (WCSP),} Oct. 2010, pp. 1-5.

\bibitem{Gui13}
G. Gui and F. Adachi, ``Improved adaptive sparse channel estimation using least mean square algorithm,'' \emph{EURASIP Journal on Wireless Communication and Networking,} vol. 2013, no. 1, pp. 1-18, 2013.

\bibitem{Gui14}
G. Gui, W. Peng, and F. Adachi, ``High-resolution compressive channel estimation for broadband wireless communication systems,'' \emph{International Journal of Communication Systems,} vol. 27, no. 10, pp. 2396-2407, Dec. 2014.

\bibitem{Ren13}
X. Ren, W. Chen, and Z. Wang, ``Low coherence compressed channel estimation for high mobility MIMO OFDM systems,'' in \emph{Global Communications Conference (GLOBECOM)}, Dec. 2013, pp. 3389-3393.

\bibitem{Ren15ICC}
X. Ren, X. Shao, M. Tao, and W. Chen, ``Compressed channel estimation for high mobility OFDM systems: pilot symbol and pilot pattern design,'' in \emph{IEEE International Conference on Communications (ICC)},    Dec. 2015, pp. 4553-4557.



\bibitem{Ren14}
X. Ren, W. Chen, and M. Tao, ``Position-based compressed channel estimation and pilot design for high mobility OFDM systems," \emph{IEEE Transactions on Vehicular Technology,} vol. 64, no. 5, pp. 1918-1929, May 2015.

\bibitem{Ren15}
X. Ren,  M. Tao, and W. Chen, ``Compressed channel estimation with position-based ICI elimination for high mobility SIMO-OFDM systems," \emph{IEEE Transactions on Vehicular Technology}, vol. 65, no. 8, pp. 6204-6216, Aug. 2016.



\bibitem{Wu2}
Q. Wu, M. Tao,  W. K. Ng, W. Chen, and R. Schober, ``Energy-efficient resource allocation for wireless powered communication networks,'' \emph{IEEE Transactions on Wireless Communications,} vol. 15, no. 3, pp. 2312-2327, Mar. 2016.



\bibitem{Wu4}
Q. Wu, W. Chen, W. K. Ng, J. Li, and R. Schober, ``User-Centric Energy Efficiency Maximization for Wireless Powered Communication Networks,'' \emph{IEEE Transactions on Wireless Communications,} vol. 15, no. 19, pp. 6898 - 6912, July. 2016.

\bibitem{Wu5}
Q. Wu, G. Y. Li, W. Chen, and W. K. Ng, ``Energy-efficient Small Cell with Spectrum Power trading,'' \emph{IEEE Journal on Selected Areas in Communications, Green communications series,} vol. 34, no. 12, pp. 3394-3408, Aug. 2016.




\bibitem{Campos13}
F. P. Campos, R. C. Alvarez, O. L. Gandara, and R. P. Michel, ``Estimation of fast time-varying channels in OFDM systems using two-dimensional prolate,'' \emph{IEEE Trans. Wireless Commun.,} vol. 12, no. 2, pp. 898-907, Feb. 2013.

\bibitem{Aboutorab12}
N. Aboutorab, W. Hardjawana, and B. Vucetic, ``A new iterative Doppler-assisted channel estimation joint with parallel ICI cancellation for high mobility MIMO-OFDM systems,'' \emph{IEEE Transactions on Vehicular Technology,} vol. 51, no. 4, pp. 1577-1589, May 2012.

\bibitem{Simon12}
E. Simon, L. Ros, H. Hijazi, and M. Ghogho, ``Joint carrier frequency offset and channel estimation for OFDM systems via the EM algorithm in the presence of very high mobility,'' \emph{IEEE Trans. Signal Processing,} vol. 60, pp. 754-765, 2012.


\bibitem{Almedia16}
J. Almeida, M. Alam, J. Ferreira and A.S.R. Oliveira, ``Mitigating adjacent channel interference in vehicular communication systems," \emph{Digital Communications and Networks}, vol.2, no. 2, pp. 57-64, May 2016.

\bibitem{Cheng13}
P. Cheng, Z. Chen, Y. Rui, Y. Guo, L. Gui, M. Tao, and Q. Zhang, ``Channel estimation for OFDM systems over doubly selective channels: a distributed compressive sensing based approach,'' \emph{IEEE Transactions on
Communications,} vol. 61, no. 10, pp. 4173-4185, Oct. 2013.

\bibitem{Tian12}
L. Tian, J. Li, J. Shi, and J. Zhou, ``Seamless dual-link handover scheme in broadband wireless communication systems for high-speed rail,'' \emph{IEEE Journal on Selected Areas in Communications}, vol. 30, no. 4, pp. 708-718,  May 2012.

\bibitem{Zhu11}
H. Zhu, ``Performance comparison between distributed antenna and microcellular systems,'' \emph{IEEE Journal on Selected Areas in Communications}, vol. 29, no. 6, pp. 1151-1163, Jun. 2011.

\bibitem{WINNER07}
K. Pekka et al., ``WINNER II Channel Models. IST-4-027756,'' \emph{Wireless World-Initiative-New-Radio (WINNER II), Munich, Germany, Tech. Rep.} D1.1.2 v1.1, Sep. 2007.

\bibitem{Liu11}
K. Liu, H. B. Lim, E. Frazzoli, H. Ji, and V. C. S. Lee, ``Improving positioning accuracy using GPS pseudorange measurements for cooperative vehicular localization,'' \emph{IEEE Transactions on Vehicular Technology}, vol.~63, no.~6, pp.~2544-2556, Jul. 2011.

\bibitem{Pascoe09}
R. Pascoe and T. Eichorn, ``What is communication-based train control?'' \emph{IEEE Vehicular Technology Magazine,} vol. 4, no. 4, pp. 16-21, Dec. 2009.


\bibitem{Hlawatsch11}
F. Hlawatsch and G. Matz, \emph{Wireless Communications over Rapidly Time-Varying Channels.} New York, NY, USA: Academic, 2011.



\bibitem{Hijazi10}
H. Hijazi and L. Ros, ``Joint data QR-detection and Kalman estimation for OFDM time-varying  Rayleigh channel complex gains,'' \emph{IEEE Trans. Commun.}, vol. 58, no. 1, pp. 170-178, Jan. 2010.


\bibitem{Kannu05}
A. Kannu and P. Schniter, ``MSE-optimal training for linear time-varying channels,'' in \emph{Proc. IEEE Int. Conf. Acoust., Speech, Singal Process. (ICASSP)}, Mar. 2005, pp. 789-792.

\bibitem{Leus04}
G. Leus, ``On the estimation of rapidly time-varying channels,'' in \emph{Euro. Signal Process. Conf. (EUSIPCO),}  Sep. 2004, pp. 2227-2230.

\bibitem{Candes06}
E. Candes, J. Romberg, and T. Tao, ``Robust uncertainty principles: exact signal reconstruction from highly incomplete frequency information,'' \emph{IEEE Trans. Inf. Theory}, vol. 52, no. 2, pp. 489-509, Feb. 2006.


\bibitem{Elad07}
M. Elad, ``Optimized projections for compressed sensing," \emph{IEEE Transcations on Signal Processing,} vol. 55, no. 12, pp. 5695-5702, Dec. 2007.

\bibitem{Hou16}
Z. Hou, Y. Zhou, L. Tian, J. Shi, Y. Li, and B. Vucetic, ``Radio environment map aided Doppler shift estimation in LTE-railway,'' to appear in \emph{IEEE Transactions
on Vehicular Technology}.

\bibitem{Chen01}
S. Chen, D. Donoho, and M. Saunders, ``Atomic decompostiion by basis pursuit," \emph{SIAM Review,} vol. 43, pp. 129-159, 2001.

\bibitem{Pati93}
Y. Pati, R. Rezaiifar, and P. Krishnaprasad, ``Orthogonal matching pursuit: recursive function approximation with applications to wavelet decomposition," in \emph{Proceedings of the 27th Annual Asilomar Conference
on Signals, Systems and Computers,}  Nov. 1993, pp. 40-44.



\end{thebibliography}
\end{document}